\documentclass[journal,twoside,web]{ieeecolor}

\usepackage{generic}

\usepackage{ntheorem}
\usepackage{hyperref}
\usepackage{textcomp}
\let\labelindent\relax
\usepackage{enumitem}
\usepackage{graphicx}
\usepackage{epstopdf}
\usepackage{float}
\usepackage[nocomma]{optidef}
\usepackage{dsfont}
\usepackage{times}
\usepackage{bm}
\usepackage{fancyhdr,graphicx,amsmath,amssymb}
\usepackage[caption=false,font=footnotesize]{subfig}
\usepackage{cite}
\usepackage{color}
\usepackage[algo2e,ruled,vlined]{algorithm2e}
\usepackage{placeins}

\usepackage{multirow}
\usepackage{makecell}
\usepackage{dblfloatfix}    

\newtheorem{theorem}{Theorem}
\newtheorem{remark}{Remark}

\newtheorem{lemma}{Lemma}

\newtheorem{proposition}{Proposition}

\newcommand{\abs}[1]{ \left| #1 \right| }
\newcommand{\inner}[1]{\left\langle #1 \right\rangle }
\newcommand{\prth}[1]{\left( #1 \right) }
\newcommand{\sqr}[1]{\left[ #1 \right] }
\newcommand{\crl}[1]{\left\{ #1 \right\} }
\newcommand{\eval}[1]{\left. #1 \right| }

\newcommand{\eps}{\varepsilon}

\newcommand{\A}{\mathcal{A}}

\newcommand{\D}{\mathcal{D}}
\newcommand{\E}{\mathcal{E}}
\newcommand{\G}{\mathcal{G}}
\newcommand{\N}{\mathcal{N}}
\renewcommand{\P}{\mathcal{P}}

\newcommand{\J}{\mathbb{J}}
\newcommand{\R}{\mathbb{R}}

\DeclareMathOperator*{\argmax}{arg\,max}

\DeclareMathOperator{\diag}{diag}

\def\BibTeX{{\rm B\kern-.05em{\sc i\kern-.025em b}\kern-.08em
    T\kern-.1667em\lower.7ex\hbox{E}\kern-.125emX}}
\begin{document}

\title{ Strategic Coalitions in Networked Contest Games } 
\author{ 
    Gilberto D\'iaz-Garc\'ia$^{*}$, 
    Francesco Bullo$^{*}$, 
    Jason R. Marden$^{*}$
    \thanks{This work was partially supported by ONR grant \#N00014-20-1-2359 and AFOSR grants \#FA9550-20-1-0054 and \#FA9550-21-1-0203. A preliminary version of the work appeared in \cite{donation:conf}.}
    \thanks{$^{*}$Department of Electrical \& Computer Engineering,University of California, Santa Barbara, Santa Barbara, CA 93106 (e-mail: gdiaz-garcia@ucsb.edu, bullo@ucsb.edu, jrmarden@ucsb.edu) } 
}

\maketitle

\begin{abstract} 
    In competitive resource allocation formulations multiple agents compete over different contests by committing their limited resources in them. For these settings, contest games offer a game-theoretic foundation to analyze how players can efficiently invest their resources. In this class of games the resulting behavior can be affected by external interactions among the players. In particular, players could be able to make coalitions that allow transferring resources among them, seeking to improve their outcomes. In this work, we study bilateral budgetary transfers in contest games played over networks. Particularly, we characterize the family of networks where there exist mutually beneficial bilateral transfer for some set of systems parameters. With this in mind, we provide sufficient conditions for the existence of mutually beneficial transfers. Moreover, we provide a constructive argument that guarantees that the benefit of making coalitions only depends on mild connectivity conditions of the graph structure. Lastly, we provide a characterization of the improvement of the utilities as a function of the transferred budget. Further, we demonstrate how gradient-based dynamics can be utilized to find desirable coalitional structures. Interestingly, our findings demonstrate that such collaborative opportunities extend well beyond the typical ``enemy-of-my-enemy'' alliances.
\end{abstract}


\section{Introduction}
\IEEEPARstart{R}{esource} allocation formulations provide a tool to optimally design how an agent should expend its limited assets. In multi-agent settings, the agents compete for a finite number of rewards based on their relative resources spent. Such competitive scenarios can effectively model political campaigns, R\&D races, advertisement expenditure competitions and military conflicts~\cite{contth,conttheory}. In addition, the inherently competitive nature of multi-agent resource allocation makes it suitable for engineering applications in security deployment along multiple infrastructures~\cite{games:sec,games:poac}, protection in cyber-physical systems~\cite{games:sensor}, multi-robot task allocation~\cite{game:robots}, among others. From a game-theoretical perspective, modeling conflicts between agents as bilateral contests provide a theoretic framework for adversarial resource allocation problems. Particularly, contest games present models where multiple players strategically compete for a finite set of items by spending their resources on them.

Contests games formulations could be cataloged by the winning rule used to determine which player obtains the items~\cite{survey}. On one hand, we could define that when any player outperforms its competitors, even with an infinitesimal margin, then it obtains the item with total certainty. Contest games with this `winner-takes-all' winning rules are defined as Colonel Blotto games~\cite{borel,gw:blotto}. On the other hand, a winning probability could be assigned to all the players based on the efforts they incur to obtain the item. The map that takes players' efforts and turn it into winning probabilities are called contest success functions~\cite{csf}. For both formulations, the resulting equilibrium behavior have been widely analyzed~\cite{blotto:eq1,ads:cont} even in scenarios with multiple contested items~\cite{blotto:eq2,multi:cont}. 

In settings with multiple items, modeling interconnections between players and the contested items is relevant to characterize the resulting behavior in presence of exogenous interactions. To represent those interactions, we consider that items of the contest game are enclosed in a network structure. For instance, we can consider that the items represent physical locations. Then, the network structure models the possible paths between the contested items that the players desire to preserve by investing resources. This setting is appropriate for security games where the players want to secure paths between locations. For this model, equilibrium payoffs and strategies have been analyzed for two-player General Lotto Games, a variation of Colonel Blotto Games~\cite{net:targets}. 

\begin{figure*}[!htb]
    \centering
    \subfloat[\label{fig:ex:3cont}]{\includegraphics[height=15ex]{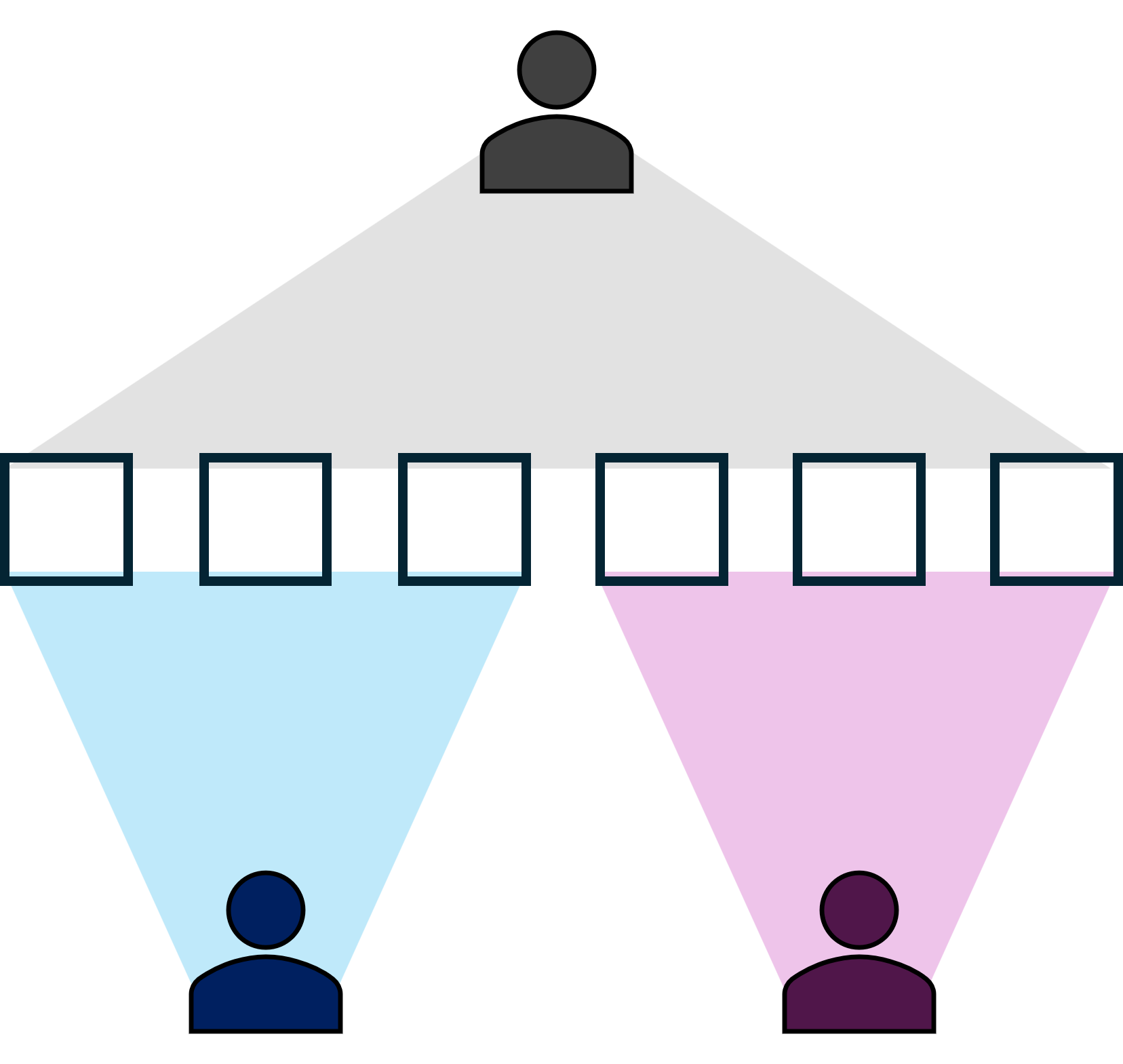}} \hfil
    \subfloat[\label{fig:ex:3graph}]{\includegraphics[height=15ex]{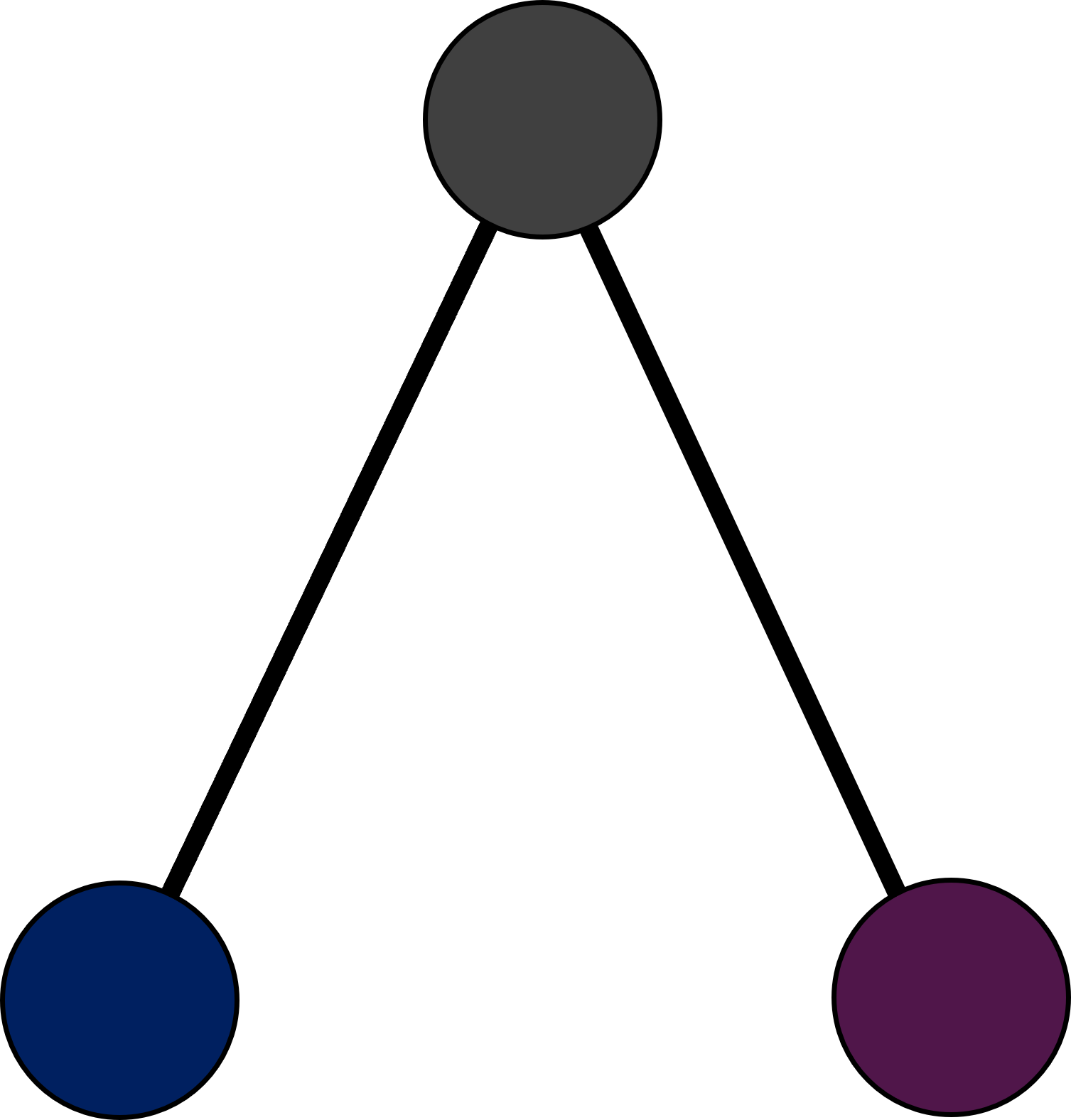} } \hfil
    \subfloat[\label{fig:ex:5cont}]{\includegraphics[height=15ex]{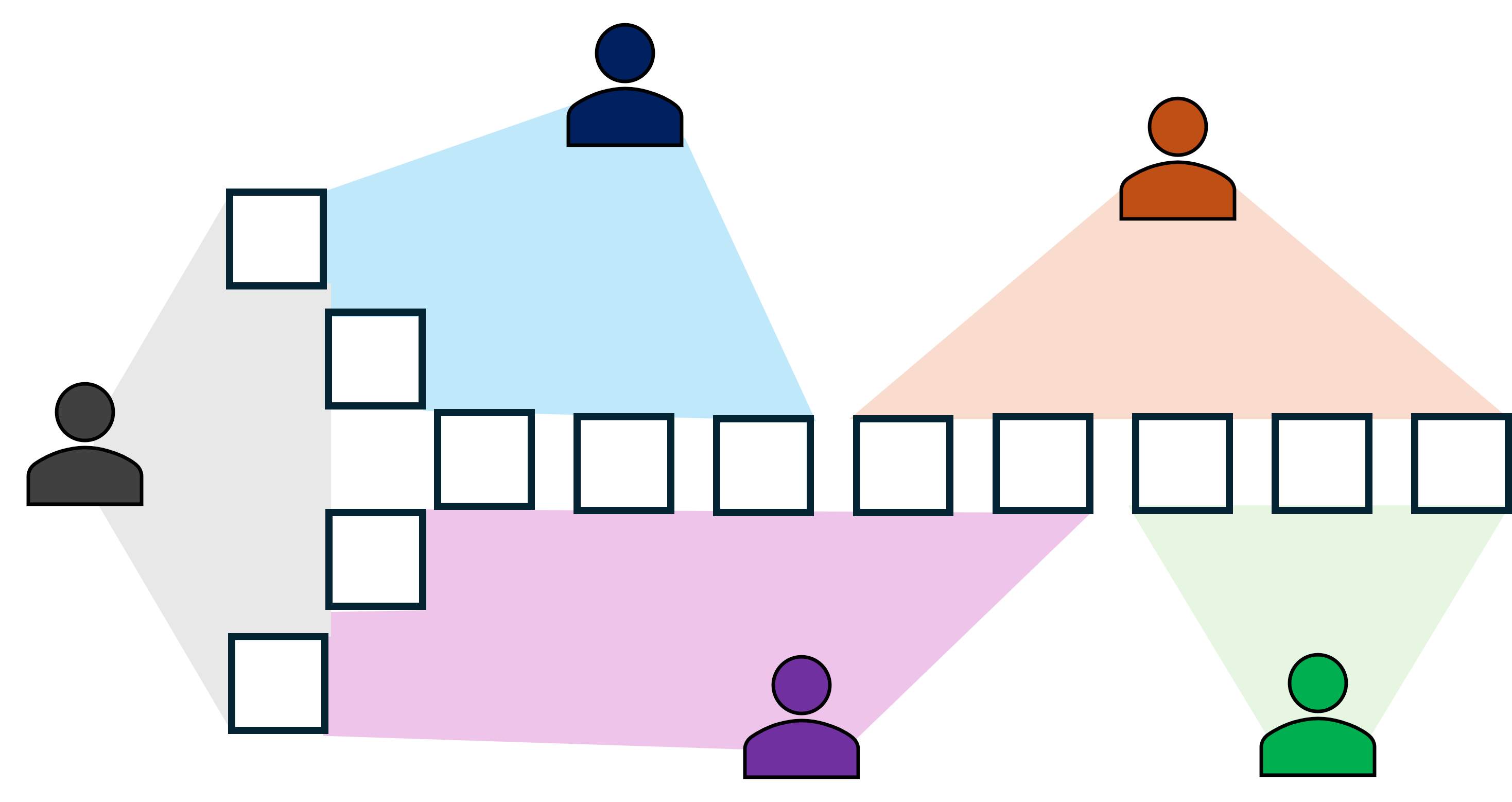} } \hfil
    \subfloat[\label{fig:ex:5graph}]{\includegraphics[height=15ex]{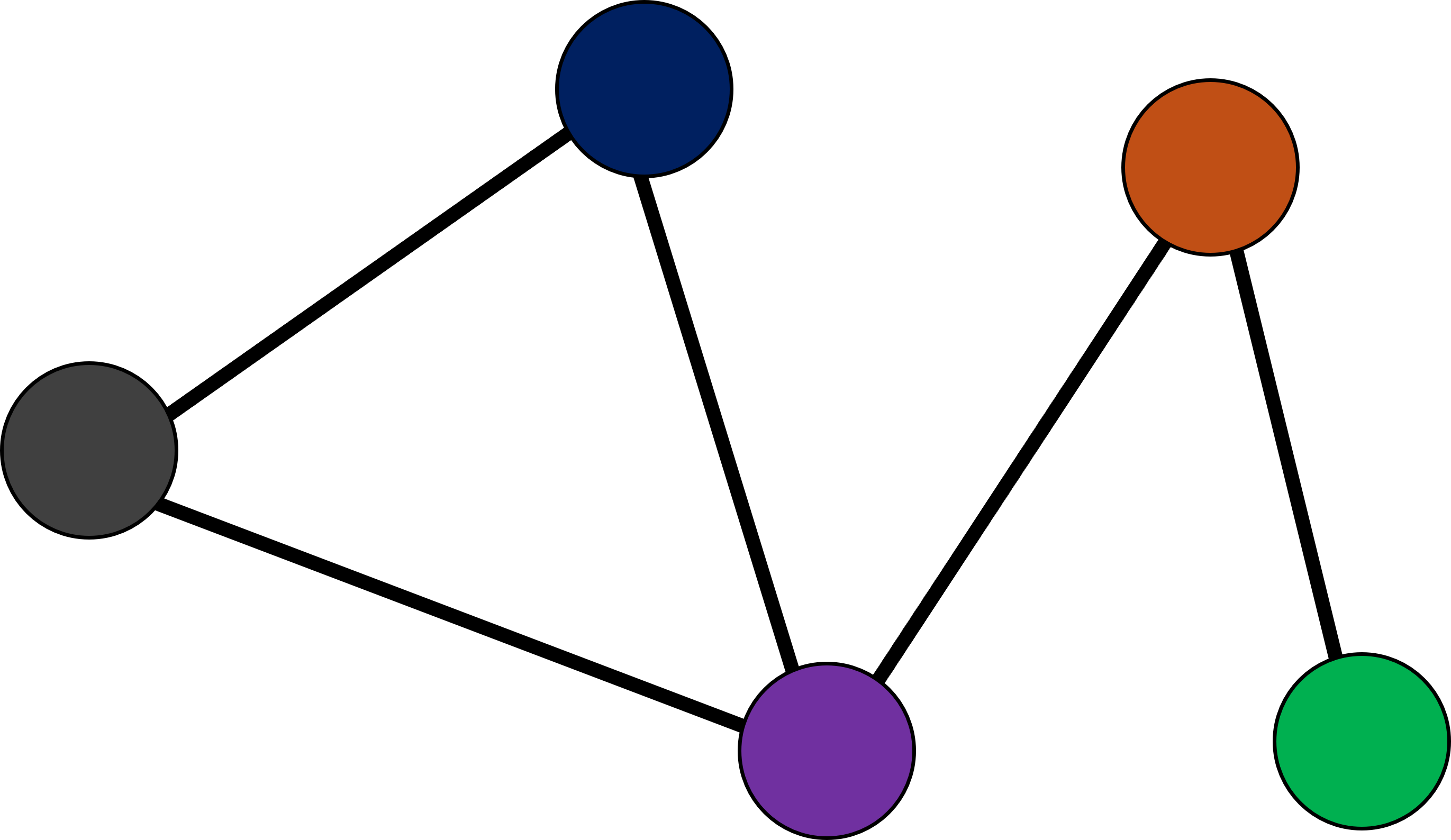} } 
    \caption{ Multiple networked contest games and their graph representation. In the contest games, it can be visualized which items (depicted as squares) each player is competing for. Equivalently, in its graph representation, an edge between two nodes indicate the existence of a bilateral contest for a common item. (a) 3-player network contest game. (b) Graph representation of the 3-player network contest game. (c) 5-player network contest game. (d) Graph representation of the 5-player network contest game. }
    \label{fig:ex:graph}
\end{figure*}

In addition, it could be possible to consider multi-player contests games. However, players may not compete for all the available items nor value them equally. These scenarios can be modeled through an undirected weighted graph where the edges represent a bilateral conflict of two players for a particular item. Therefore, there exist a direct correlation between contest games with multiple players and the graph representation of it networks structure, as illustrated in {Figure~\ref{fig:ex:graph}}. Existence and uniqueness of Nash equilibrium for this networked model have been presented for a family of winning rules similar to the Tullock contest success function~\cite{net:conf}.

To manipulate the resulting behavior in contest games, strategic opportunities can be offered to the players such as changes in the revealed information~\cite{reveal:1,reveal:2}, division of the players' assets~\cite{division} or alliances with budgetary transfers~\cite{transf:1}. In these alliances, players are able to transfer a portion of their assets to another player seeking to increase their own payoffs. Intuitively, when a player gives away resources to others players reduce its capability to compete over all the contested items. Thus, making budgetary transfers can be considered a harmful decision. However, in scenarios similar to ones depicted in Figure~\ref{fig:ex:3cont}, it has been verified that it could potentially improve the players' payoffs by making its enemies weaker. 

Budgetary transfers in the scenario depicted in Figure~\ref{fig:ex:3cont} can be thought as a coalition formation in order to overcome a mutual adversary. This idea of coalitional Blotto Games have been studied for this simplified scenario under the name of ``enemy-of-my-enemy'' alliances~\cite{transf:2}. However, as the number players are able to participate in a contest game increases, the interactions between player get more complex as illustrated in Figure~\ref{fig:ex:5cont}. Therefore, networked contest games present a natural extension of coalitions in contest games. Although, as far as we know, there is no algorithms to find beneficial alliances in general networked contest games. Moreover, given the complexity that emerges with general networks, is not entirely clear if such beneficial coalitions even exist. 

In this work, we analyze alliances with budgetary transfers for network contest games. In particular, we tackle the questions: \textit{under which network structures do there exists a mutually beneficial coalition?} and \textit{which players are worth making an alliance with?} With this in mind, we list the main contributions of this document as follows. First, we present the model for budget-constrained network contest games. Second, we provide an equivalent formulation using per unit cost for the assets that allow us to define the equilibrium strategies for all players. With them, we can recover the equilibrium payoffs for the budget constrained formulation and the equivalent budgets for the players. Then, we present sufficient conditions to ensure the existence of a transfer that benefits both players in the coalition. Moreover, we offer a construction for any contest game that ensure the existence of such transfer with any player under mild conditions on the connectivity of the conflict network. Therefore, we can assert that the existence of beneficial coalitions is independent of the network structure of the contest game and the chosen players that decide to be part of the alliance. Finally, we provide an expression for the gradient of the equilibrium payoffs as a function of the budgetary transfer between players. With it, different Nash-seeking dynamics, e.g., discrete replicator dynamics, could be implemented to define budgetary alliances that optimally improve their utilities.

The remainder of the paper is organized as follows. In Section~\ref{sec:model} we present the model for networked contest game. In Section~\ref{sec:3node} we characterize beneficial alliances in a simplified 3-node scenario where two players compete with a common enemy. Section~\ref{sec:transf} present conditions to assert the existence of a mutually beneficial transfer independently of the network structure of the game. The formulation for the gradient of the players' utilities as a function of their donated budget is introduced in Section~\ref{sec:grad}. Then, in Section~\ref{sec:sims}, we show numerical simulations to verify existence of beneficial transfers. Finally, in Section~\ref{sec:conc}, we present conclusions and future research directions. The proofs of our results are provided in the Appendix.

\section{Model}\label{sec:model}
\subsection{Contest Games in Networks}\label{sec:model:cont}
In a networked contest game a set of players $\P=\crl{1,\cdots,n}$ compete over bilateral contests to obtain different items. Each contested item have a potential benefit for the players competing for it and players compete for them by spending resources from their limited budget $B_i > 0$ for all $i\in\P$. We condense the structure of a networked contest game using an undirected graph $\G=(\P,\E,v)$. Here, the node set $\P$ represent the players and the edge set $\E \subset \P \times \P$ the contested items. For each item $(i,j) \in \E$ we define $v_{i,j} = v_{j,i} \geq 0$ as the value associated to it, representing how profitable it is.  Under this setting, the set of available actions for each player is defined as, \[ \A_i := \crl{ x_i \in \R_{\geq0}^{n_i} ~:~ \sum_{j\in\N_i} x_{i,j} = B_i }, \] where $\N_i := \crl{ j ~:~ v_{i,j} > 0 }$ is the neighbors set of the player $i$ and $n_i = \abs{\N_i}$. Thus, $x_i = \sqr{x_{i,1}, \cdots, x_{i,n} } \in \A_i$ denotes a possible allocation for player $i$ among all its contested items given that the total spent resources can not exceed its budget $B_i$. For any strategy profile $x = [x_1,\cdots,x_n]$ with $x_i \in \A_i$ the player $i$'s payoff is given by, 
\begin{equation} \label{eq:ui}
    U_i(x) = \sum_{j\in\N_i} v_{i,j} \alpha\prth{ x_{i,j} , x_{j,i} },   
\end{equation}
where $x_{i,j}$ defines the resources spent by player $i$ in the item contested with player $j$ and $\alpha\prth{ x_{i,j} , x_{j,i} } \in [0,1]$ defines the probability of player $i$ to win the item $(i,j)$. Therefore, the utility in Equation~\eqref{eq:ui} can be interpreted as the expected value obtained along all contested items for player $i$ given all players' efforts $x$. 

For these games, we are interested in the emergent behavior in competitive environments. In particular, we consider strategies that are stable under other player's decisions. For this, we focus our analysis on the Nash equilibrium strategies. Those are strategy profiles $x^* = \sqr{ x_1^*, \cdots, x_n^* }$ such that, 
\[ U_i( x_i^*, x_{-i}^* ) \geq U_i( x_i, x_{-i}^* )  ~~\forall x_i \in \A_i \text{ and } i \in \P, \]
where $x_{-i} = \prth{ x_1, \cdots, x_{i-1}, x_{i+1}, \cdots, x_n }$. This means that there is no incentive for any player to unilaterally deviate from $x^*$. Note that the emerging behavior, characterized by the equilibrium allocation $x^*$, depends on the parameters of the contest game: the budgets of the players $B_i$ for $i\in\P$ and the values of the items $v_{i,j}$ for $(i,j)\in\E$. 

In this work we focus on the ratio winning rules axiomatized in~\cite{csf}. This rules are described as, 
\begin{equation} \label{eq:tullock}
    \alpha\prth{ x_{i,j} , x_{j,i} } = \frac{ f\prth{ x_{i,j} } }{ f\prth{ x_{i,j} } + f\prth{ x_{j,i} } },
\end{equation}
where $f(x) = \mu x ^r$ with $\mu, r > 0$. In particular, the presented results use the Tullock lottery contest success function, i.e., a ratio winning rule with $\mu = 1$ and $r=1$. To avoid an ill definition of the winning rule in Equation~\eqref{eq:tullock} we need to extend its definition using a tie-braking rule such as $\alpha(0,0) \in (0,1)$. However, this extended definition does not change the results presented in this paper. 

\subsection{Alliances with Budgetary Transfers} \label{sec:model:alli}
Given the dependence of equilibrium payoff on the parameters of the game, the players may consider to make coalitions to affect the resulting equilibrium strategies. In particular, players could manipulate the budgets through coalitions with budgetary transfers. That is, for $(a,b)\in\P^2$ we can define a $\tau \in [0,B_a)$ such that $\tilde{B}_a = B_a - \tau$ and $\tilde{B}_b = B_b + \tau$. With the new set of budgets $\tilde{B}$, a new equilibrium allocation $\tilde{x}^*$ will emerge, potentially affecting the obtained payoffs for all the players in the network. We say that a transfer $\tau$ is \textit{mutually beneficial} if the new equilibrium payoff is better than the originally obtained with budgets $B$ for players $a$ and $b$, i.e., $U_a\prth{ \tilde{x}^* } > U_a( x^* )$ and $U_b\prth{ \tilde{x}^* } > U_b( x^* )$.

Intuitively, an increased budget should be beneficial to the receiving player. However, it is not clear if the payoff of the player who gives resources also improves. With this in mind, let us present the following numerical example with a mutually beneficial transfer. 

\begin{figure}[hbt]
    \centering
    \subfloat[\label{fig:ex1}]{\includegraphics[width=0.2\textwidth]{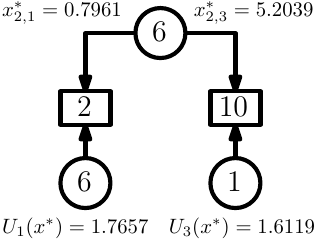}} \hfil
    \subfloat[\label{fig:ex2}]{\includegraphics[width=0.2\textwidth]{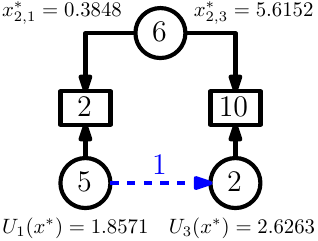}} 
    \caption{ Numerical Example of Contest Game in a Network. (a) Without transfer between players. (b) With a mutually beneficial transfer. }
    \label{fig:ex}
\end{figure}

Let us consider the 3-player contest game with $\sqr{B_1,B_2,B_3} = \sqr{6,6,1}$, $v_{1,2}=2$ and $v_{2,3} = 10$ as shown in Figure~\ref{fig:ex1}. With these values of $B$ and $v$ the players can define their equilibrium strategy $x^*$ and receive their corresponding payoff $U_1(x^*)\approx1.7657$ and $U_3(x^*)\approx1.6119$. Now, consider the budget transfer $\tau=1$ from player $1$ to player $3$, as in Figure~\ref{fig:ex2}. As expected, player $3$ significantly increases its equilibrium payoff since it double its budget, obtaining a payoff $U_3(x^*)\approx2.6263$. More importantly, the increased budget of player $3$ alters others players behavior. From this change, player $1$ also receives a higher payoff $U_1(x^*)\approx1.8571$. Therefore, there exists a mutually beneficial transfer for the setting in Figure~\ref{fig:ex}.

This presented example suggests there exists cases when giving some assets to another player could potentially increase the giving player's payoff. Therefore, two players can be motivated to form a coalition that allows them to transfer resources between them. However, it is not known if situations where beneficial transfers for both players will ever present themselves. With this in mind, we devote the rest of the paper to determine if there exists items valuations $v_{i,j}$ and budgets $B_i$ for a given networked contest game $\G$ such that there is a mutually beneficial transfer between players $a$ and $b$.

\begin{figure*}[!htb]
    \centering
    \subfloat[\label{fig:graph:orig}]{\includegraphics[width=0.2\textwidth]{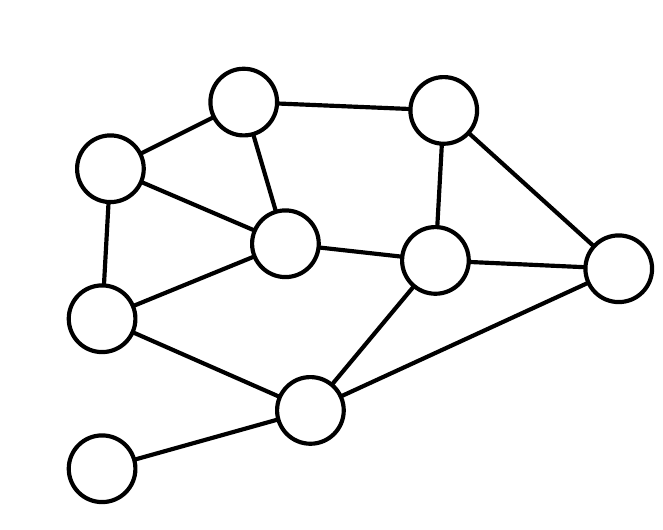}} \hfil
    \subfloat[\label{fig:graph:line}]{\includegraphics[width=0.2\textwidth]{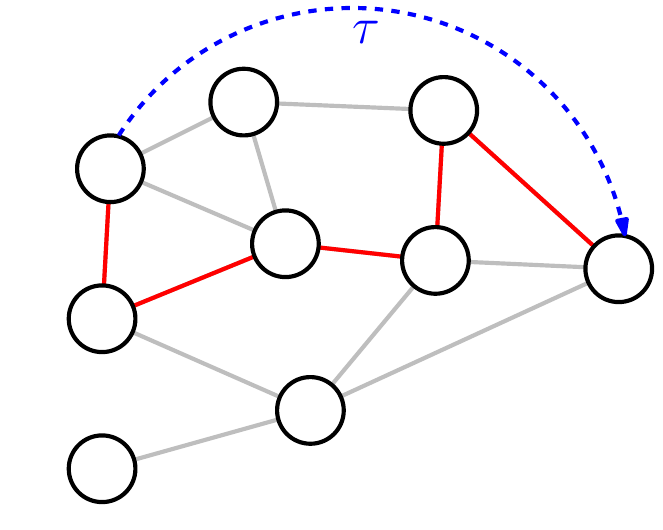} } \hfil
    \subfloat[\label{fig:graph:cycl}]{\includegraphics[width=0.2\textwidth]{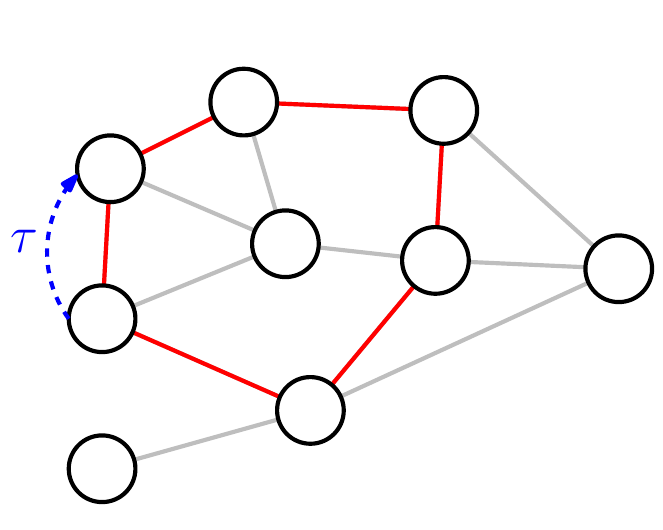} } \hfil
    \subfloat[\label{fig:graph:neig}]{\includegraphics[width=0.2\textwidth]{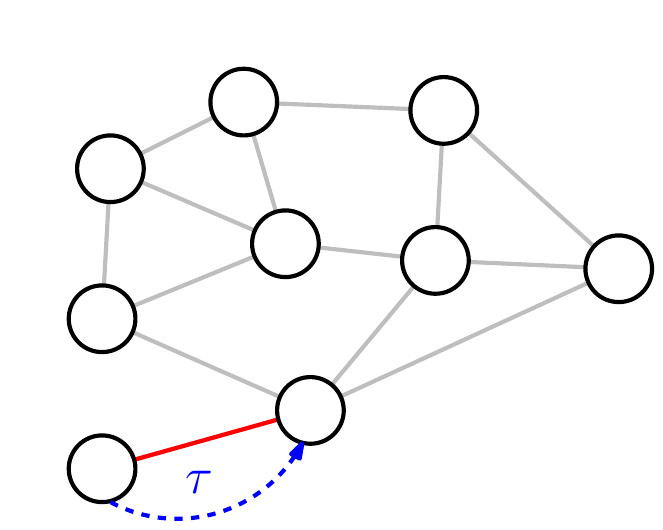} } 
    \caption{ Existence results for donations in networked contest games. (a) Original conflict graph. (b) Donations between players that are not neighbors. (c) Donations between players that are neighbors. (d) Donations from a player to its only neighbor.  }
    \label{fig:graph}
\end{figure*}

\section{Existence of Mutually Beneficial Transfer in Networked Contest Games}\label{sec:transf}
In this section, we analyze how budgetary transfers between two players are potentially beneficial in settings involving complex networks. Moreover, we devote this section to ensure that the existence of a mutually beneficial transfer is independent of the conflict network $\G = (\P,\E,v)$ that support the contest game. This means that the strategic opportunities to increase the payoff for a particular player through alliances are not limited to its local network. With that in mind, let us present our main result as follows, 
\begin{theorem}~\label{thm:graph}
    Consider any graph $\G=(\P,\E,v)$ and any pair of players $(a,b)\in\P^2$. If there is a path between players $a$ and $b$ in the subgraph $\G_{a,b}=(\P,\E_{a,b},v)$, with $\E_{a,b} = \E \setminus \crl{(a,b)}$, then there exists an instance of a networked contest game such that transferring resources from player $a$ to player $b$ is mutually beneficial. 
\end{theorem}
With Theorem~\ref{thm:graph} we can guarantee that, independently of the conflict network $\G$, there exists items valuations $v_{i,j}$ for $(i,j)\in\E$ and player budgets $B_i$ for $i\in\P$ such that transferring resources from player $a$ to player $b$ improves the obtained utility for both players. 

To verify Theorem~\ref{thm:graph} we start by characterizing the parameter space, budgets $B$ and values $v$, that allows the existence of a mutually beneficial transfers between two players that compete against a common enemy. These alliances are known as \textit{`enemy-of-my-enemy' alliances} and have been analyzed using a different model of contest games~ \cite{transf:1,transf:2}. 

As we move to scenarios with complex networks, we identify conditions to ensure that the equilibrium payoff of any player strictly increases for a given transfer $\tau > 0$ from player $a$ to player $b$. Then, for any given graph, we can build an instance of a networked contest game where that transfer is mutually beneficial for player $a$ and $b$. With this in mind, we aim to prove three existence results, \begin{itemize}
    \item existence of mutually beneficial transfers in line graphs (Figure~\ref{fig:graph:line}),
    \item existence of mutually beneficial transfers in cycle graphs  (Figure~\ref{fig:graph:cycl}) and, 
    \item inexistence of mutually beneficial transfers from a player to its only direct competitor  (Figure~\ref{fig:graph:neig}).
\end{itemize}
These partial results provide tools to build instances of mutually beneficial transfers for any given graph $\G$. 

\subsection{ Characterization of `Enemy-of-my-Enemy' Alliances }\label{sec:3node}
\begin{figure}[htb!]
    \centering
    \includegraphics[width=0.15\textwidth]{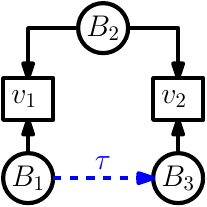}
    \caption{Graph representation of a `Enemy-of-my-Enemy' Alliance. In these alliances players $1$ and $3$ make a coalition to outperform their common adversary, player $2$.}
    \label{fig:3node}
\end{figure}

The idea of alliances with budgetary transfers in contest games have been previously explored in \cite{transf:1,transf:2}. In their works, the authors analyze the individual incentives of two players to form an alliance, allowing them to transfer resources, while competing with a common rival as depicted in Figure~\ref{fig:3node}. While the items are modeled as disjoints Colonel Blotto games, which use different winning rules as those presented in Equation~\eqref{eq:tullock}, their notion of alliances share some similarities with the example presented in Section~\ref{sec:model:alli}. Particularly, in their characterization of 'enemy-of-my-enemy' alliances, the players can be encouraged to share resources between them depending on the budgets of the players and the values of the items. This dependence on the game parameters is a crucial factor that will be used through our discussion. 

In general, `enemy-of-my-enemy' alliances are a simplified version of alliances in networked contest games. However, they provide an illustration of how players could be incentivized to form alliances with budgetary transfers. Thus, let us introduce a full characterization of the scenarios where players will form an alliance under the model presented in Section~\ref{sec:model:cont}. 
\begin{proposition} \label{prop:3node}
    For a 3-node contest game, as presented in Figure~\ref{fig:3node}, there is a mutually beneficial transfer from player $1$ to player $3$ if, 
    \begin{equation*}
        \frac{(B_1 + B_2)^2}{B_1B_3} > \frac{v_1}{v_2} > \frac{B_1 B_3}{ (B_2 + B_3)^2 }
    \end{equation*}
    and,
    \begin{equation*}
        B_1 - B_3 > 2\sqrt{ \frac{v_1}{v_2} B_1 B_3 }.
    \end{equation*}
\end{proposition}

\begin{figure}[hbt!]
    \centering
    \subfloat[$v_1/v_2 = 0.1$ \label{fig:reg:1}]{\includegraphics[width=0.23\textwidth]{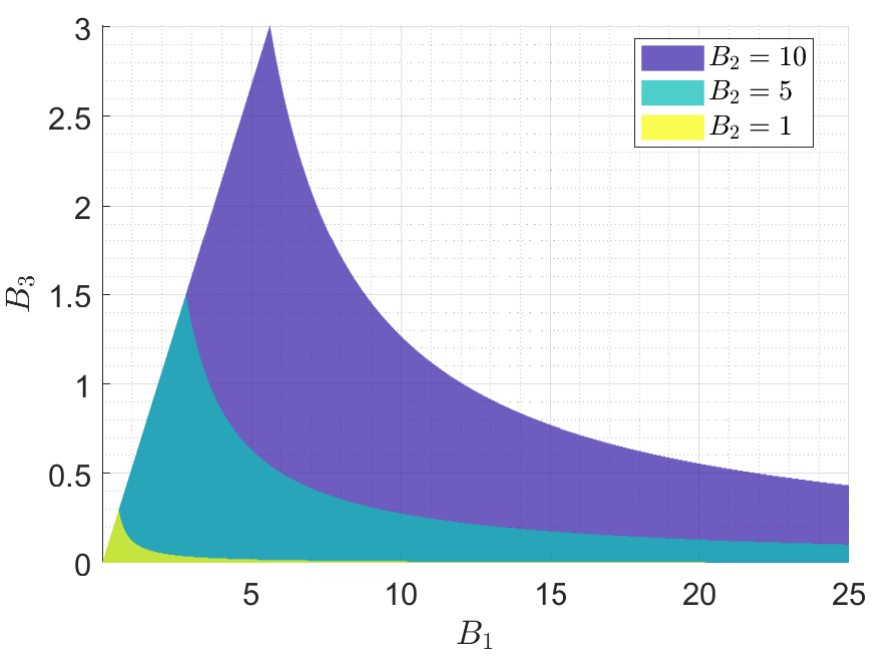}} \hfil
    \subfloat[$v_1/v_2 = 0.5$ \label{fig:reg:2}]{\includegraphics[width=0.23\textwidth]{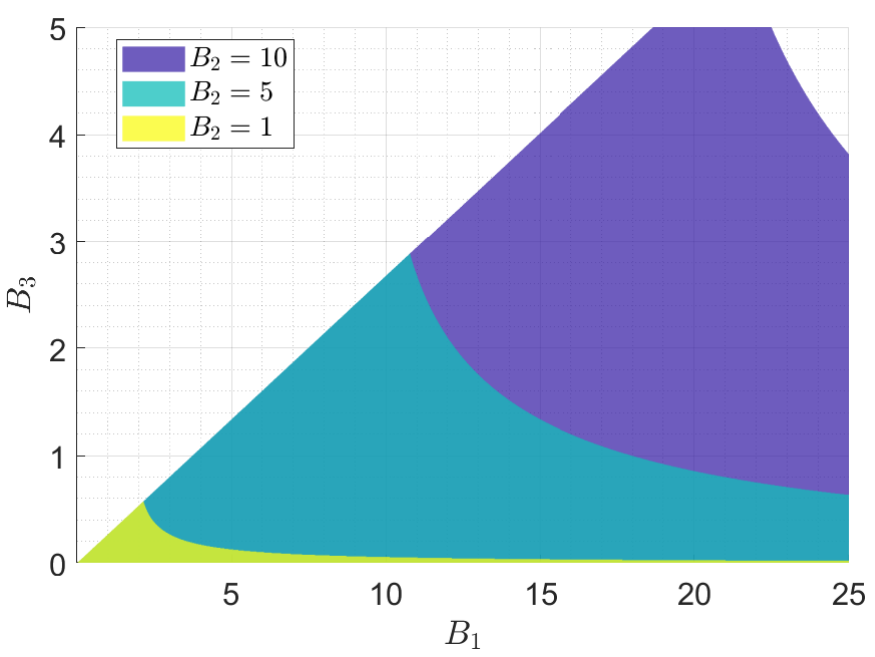}}  \\
    \subfloat[$v_1/v_2 = 1$   \label{fig:reg:3}]{\includegraphics[width=0.23\textwidth]{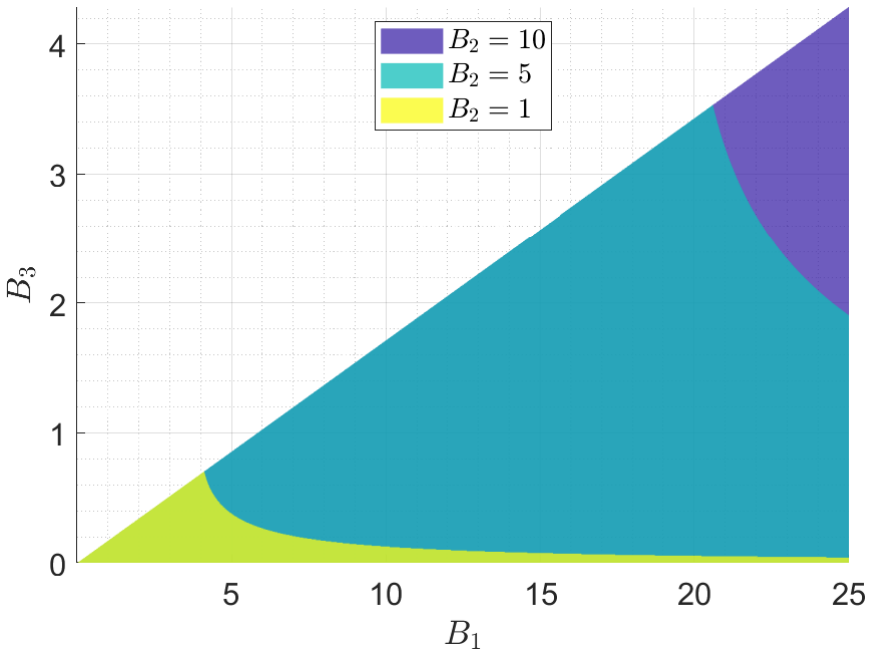}}  
    \caption{ Regions with mutually beneficial `enemy-of-my-enemy' alliances. Here, any set of parameters $B_1$, $B_2$, $B_3$ and $v_1/v_2$ inside of the shaded regions verify the existence of a mutually beneficial transfer between players $1$ and $3$. }
    \label{fig:reg}
\end{figure}

Note that the values of the items always appear as the ratio $v_1/v_2$. Then, Proposition~\ref{prop:3node} offer a full characterization where the exists a mutually beneficial donation over the $4$-dimensional space $\sqr{B_1,B_2,B_3,v_1/v_2}$. In Figure~\ref{fig:reg} this region is depicted after fixing multiple values of $B_2$ and $v_1/v_2$. 

Regions illustrated in Figure~\ref{fig:reg} characterized the set of parameters $B_1$, $B_2$, $B_3$ and $v_1/v_2$ such that there exist some value $\tau > 0$ that makes the utilities for players $1$ and $3$ increase after define the new budgets $\tilde{B}_1 = B_1 - \tau$ and $\tilde{B}_3 = B_3 + \tau$. Alternatively, if the parameters of the contest game do not lie inside those regions, then there are no strategic opportunities for the players to improve their payoffs via budgetary transfers.

\subsection{ Per-unit-cost Parametrization of Contest Games }
As the complexity of the conflict network $\G$ increases computing the equilibrium payoffs quickly becomes a daunting task. To characterize the equilibrium strategies of the players in a networked contest game all the possible interactions between them must be considered. However, these dependencies can be decoupled through an equivalent reformulation of the contest game. Let us start by defining the best response for each player, 
\[ BR\prth{x_{-i}} := \argmax_{ x_i \in \A_i } ~~ U_i( x_i, x_{-i} ). \]

Note that the equality constraint included in $\A_i$ could be removed through the use of a Lagrange multiplier $\lambda_i$ as follows, 
\begin{align*}
    BR\prth{x_{-i}}     &= \argmax_{ x_i \in \A_i } ~~ U_i( x_i, x_{-i} ), \\
                        &= \argmax_{ x_i \in R_{\geq0}^{n_i} } ~~ U_i( x_i, x_{-i} ) - \lambda_i \sum_{j\in\N_i} x_{i,j},
\end{align*}
for some $\lambda_i\in\R$ for each $i\in\P$. Therefore, we can reformulate the networked contest game in a setting where the players, instead of having a limited budget $B_i$, have an unlimited budget but they are charged $\lambda_i > 0$ per unit of resources spent. With this in mind, let us define the alternative payoffs, \begin{equation}\label{eq:Uih}
    \hat{U}_i( x_i, x_{-i} ) = U_i( x_i, x_{-i} ) - \lambda_i \sum_{j\in\N_i} x_{i,j}.
\end{equation}
where $\lambda_i$ is the cost of allocating resources for player $i$. We called the formulation with the payoffs in Equation~\eqref{eq:Uih} the \textit{per-unit-cost parametrization} of the networked contest game. While both formulations have the same best response, for the per-unit-cost parameterization we can find the Nash equilibrium, the equilibrium payoff and the equivalent budget for each player.

\begin{lemma}\label{lm:eqCost}
    For the per-unit-cost parametrization of the networked contest game with costs $\lambda_i$ and payoffs defined as in Equation~\eqref{eq:Uih} the Nash equilibrium is described by the allocations, \[ x_{i,j}^* = v_{i,j} \frac{\lambda_j}{ \prth{ \lambda_i + \lambda_j }^2 }. \]
    
    Therefore, the equilibrium payoffs for the equivalent budget constrained contest game are, 
    \begin{equation}\label{eq:Ui*}
        U_i^* := U_i( x^* ) = \sum_{j\in\N_i} v_{i,j} \frac{\lambda_j}{\lambda_i+\lambda_j}.
    \end{equation}
    
    Moreover, any instance of a per-unit-cost network contest game is equivalent to budget-constrained network contest game with budgets,   
    \begin{equation}\label{eq:Bi}
        B_i\prth{\lambda_1,\cdots,\lambda_n} = \sum_{j\in\N_i}  v_{i,j} \frac{\lambda_j}{ \prth{ \lambda_i + \lambda_j }^2 }.
    \end{equation}
\end{lemma}

Lemma~\ref{lm:eqCost} presents two important results. First, we obtain an entire characterization of the Nash equilibrium for per-unit-cost contest games. Additionally, we establish the equivalence between budget constrained contest games and its per-unit-cost parametrization. Therefore, without losing generality, we can analyze networked contest games from its budget constrained formulation or its corresponding per-unit-cost parametrization. 

\subsection{ Overview of Proof of Theorem~\ref{thm:graph} }
Now, we devote the rest of the section to provide a proof for Theorem~\ref{thm:graph}. As it was previously mentioned, we start by characterizing necessary conditions to the existence of a mutually beneficial transfer. Then, we prove the existence of mutually beneficial transfers for the particular cases shown in Figure~\ref{fig:graph}. With them, we are able to construct an instance with a mutually beneficial transfer under mild connectivity conditions. 

\subsubsection{Sufficient Conditions for a Beneficial Transfer}
Let us assume that there is budgetary transfer $\tau > 0$ from player $a$ to player $b$. Then, the budgets of the players are redefined as $\tilde{B_a} = B_a - \tau$ and $\tilde{B_b} = B_b + \tau$. In vectorial form we have that $\tilde{B} = B + \tau \prth{ e_b - e_a }$ with $e_i$ the $i$th vector of the canonical base. Note that, 
\begin{equation*} 
    \frac{\partial \tilde{B} }{\partial \tau} =  e_b - e_a,
\end{equation*}
particularly at $\tau = 0$. Using this, we can state a sufficient condition on the valuations and per unit cost that guarantee the existence of a positive transfer as follows, 
\begin{lemma}~\label{lm:cond}
    For any player $i\in\P$, there exists a budget transfer $\tau > 0$ that is beneficial to player $i$ if,
    \begin{equation}\label{eq:cond}
        \frac{\partial U_i^*}{\partial \tau} = \nabla_\lambda^\top U_i^* ~~ \J^{-1}_{B/\lambda}(\lambda) ~~ \frac{\partial \tilde{B} }{\partial \tau} > 0
    \end{equation} 
    where $\nabla_\lambda U_i^*$ is the gradient of $U_i^*$, as in Equation~\eqref{eq:Ui*}, with respect to $\lambda$ and $\J_{B/\lambda}(\lambda)$ is the Jacobian matrix of the map  $B : \R^n_{>0} \to \R^n_{>0}$ defined in Equation~\eqref{eq:Bi}.
\end{lemma}
\begin{remark}~\label{rm:cond}
    Since $\frac{\partial \tilde{B} }{\partial \tau} = e_b - e_a$ then, condition in Equation~\eqref{eq:cond} can be written as $q_b > q_a$ where $q_k$ is the $k$th entry of the vector that solves the equation $\J_{B/\lambda}^\top(\lambda) q = \nabla_\lambda U_i^*$. 
\end{remark}

Using the condition presented in Lemma~\ref{lm:cond} we can build an instance where a mutually beneficial transfer exists between players $a$ and $b$ for an arbitrary graph that remains connected after removing the edge $(a,b)$. 

\subsubsection{Mutually Beneficial Transfers in Line Graphs}
Now, let us assume that the players $a$ and $b$ are not competing for any items, i.e., $(a,b) \notin \E$. In addition, recall that there is at least one path that connects $a$ and $b$ in the graph $\G = \G_{a,b}$. Therefore, to guarantee the existence instance with a mutually beneficial transfer is sufficient to build an instance for the line graph that connects them. For line graphs, we can state the following inductive argument. 
\begin{lemma}~\label{lm:line}
    If there is an instance of a mutually beneficial transfer for a $(n-2)$-line graph, then there exist at least one instance for a $n$-line graph. 
\end{lemma}

Lemma~\ref{lm:line} asserts that, if two players can make a mutually beneficial alliance along a line graph then, two additional players can be added between them and maintain the benefit of the alliance. Therefore, we can generalize the result for line graph with arbitrary size. 
\begin{remark}~\label{rm:line}
    If there is at least one instance for a $3$-line graph and for a $4$-line graph, then there is at least one instance for a $n$-line graph with $n>2$. 
\end{remark}

With Remark~\ref{rm:line} we can guarantee the existence of at least one instance of a mutually beneficial transfer between any two players that are not neighbors in the conflict network $\G$. 

\subsubsection{Mutually Beneficial Transfers in Cycle Graphs}
From the statement in Lemma~\ref{lm:line} seems reasonable to think that alliances between players that are competitors. However, the possible drawback of donating a direct enemy can be neutralized by the attention it can draw from other competitors. For instance, in cycle graphs, a player could donate assets to one of its neighbors if it makes easier to compete with its other neighbor. Thus, we use cycle graphs as a foundational cases where players can donate to its neighbors. 
\begin{lemma}~\label{lm:cycle}
    If there is an instance of a mutually beneficial transfer for a $n$-line graph then, there exist at least one instance for a $n$-cycle graph. 
\end{lemma}

Using the results of Lemma~\ref{lm:cycle} and Remark~\ref{rm:line} assure the existence of a mutually beneficial alliance between players that are competing in the conflict network $\G$. 

\subsubsection{Mutually Beneficial Transfers to a Unique Competitor}
Finally, we analyze the cases where a player would like to donate to its only neighbor. This will provide a full characterization of the set of players which are worth to make an alliance with. Note that giving assets to your only competitor intuitively can only make your performance worse. However, in more complex networks, it is not entirely clear if that will always be the case. With this in mind, let us assert the following result. 
\begin{lemma}~\label{lm:neigh}
    For any graph $\G$ and any pair of players $(a,b)\in\P^2$ such that $\N_a = \crl{b}$ there is no instance where a mutually beneficial transfer exist from player $a$ to player $b$.
\end{lemma}

Hence, using results in Lemmas~\ref{lm:line},~\ref{lm:cycle} and~\ref{lm:neigh}, we can assert result in Theorem~\ref{thm:graph}. This means that alliances, in terms of budgetary transfers, represent strategic opportunities for every player to improve their payoff. While the existence of mutually beneficial transfers requires mild connectivity conditions on the conflict graph $\G$, roughly such existence is independent of the conflict graph and the chosen player to make the coalition. 

\section{Optimally Designed Budgetary Transfers}\label{sec:grad}
Now that the existence of potentially mutually beneficial transfers has been established, we will focus on how to design a transfer in a particular instance of a networked contest game. With that in mind, in this section we describe how the equilibrium payoffs $U_i^*$ vary as a function of the donations that players can make between them. 

First, let us define the undirected donations graph $\G_D = \prth{\P,\E_D}$. Here, $(a,b)\in \E_D$ means that player $a$ have the option to donate to player $b$. Similarly to the conflict graph, let us define the set of potential allies of player $i$ as $\D_i =  \crl{ j ~:~ (i,j) \in \E_D } \cup \crl{i}$ and $d_i = \abs{\D_i}$. Then, player $i$ could define its vector of donations $b_i = \sqr{b_{i,1} , \cdots , b_{i,n} } \in \R_{+}^n$ such that $\sum_j b_{i,j} = 1$. Hence, $b_{i,j}$ represent the fraction of the budget donated from player $i$ to player $j$. Note that this formulation captures scenarios without donations by fixing $b_{i,i} = 1$ for all $i\in\P$. Using the definition of $b_i$, we can redefine the budgets of the players after the transfer as, \[ \tilde{B}_i = \sum_{ j \in \D_i } b_{j,i}B_j. \]

Then, using the equivalence between budgets and per-unit costs in Equation~\eqref{eq:Bi} and the definition of equilibrium payoffs in Equation~\eqref{eq:Ui*} we can find the gradient of $U_i^*$ with respect to the donations vectors $b_i$ for all $i \in\P$. 
\begin{proposition} \label{prop:grad}
    The gradient of the equilibrium payoffs $U_i^*$ with respect to the donation vector $b_i$ is, 
    \begin{equation} \label{eq:grad}
        \nabla_{b_i} U_i^* = B_i \sqr{ \J^{\top}_{\tilde{B}/\lambda}(\lambda) }^{-1} \nabla_\lambda U_i^*, 
    \end{equation}
    where $\lambda$ satisfies the equation, 
    \begin{equation*}
        \sum_{ j \in \D_i } b_{j,i}B_j = \tilde{B}_i = \sum_{j\in\N_i}  v_{i,j} \frac{\lambda_j}{ \prth{ \lambda_i + \lambda_j }^2 },
    \end{equation*}
    for all $i\in\P$. 
\end{proposition}

With a expression for $\nabla_{b_i} U_i^*$ players can design their donations in order to maximize $U_i^*$. With this in mind, gradient-based Nash seeking algorithms could be used in order to design the donation vectors $b_i$. 

\section{Simulations}\label{sec:sims}
In this section, we present numerical simulations to highlight the behavior presented in Sections~\ref{sec:transf} and~\ref{sec:grad}. First, let us revisit the scenario presented in Figure~\ref{fig:ex}. Using the result in Proposition~\ref{prop:3node}, we can verify if for that set of parameters a mutually beneficial transfer exists. Checking the two conditions we obtain that, 
\begin{align*}
    \frac{(B_1 + B_2)^2}{B_1B_3} > \frac{v_1}{v_2} > \frac{B_1 B_3}{ (B_2 + B_3)^2 } & \iff 24 > \frac{1}{5} > \frac{6}{49}, \\
    B_1 - B_3 > 2\sqrt{ \frac{v_1}{v_2} B_1 B_3 } & \iff 5 > 2 \sqrt{ \frac{6}{5} }.
\end{align*}

\begin{figure}[hbt]
    \centering
    \subfloat[Player $1$ \label{fig:3node:a}]{\includegraphics[width=0.24\textwidth]{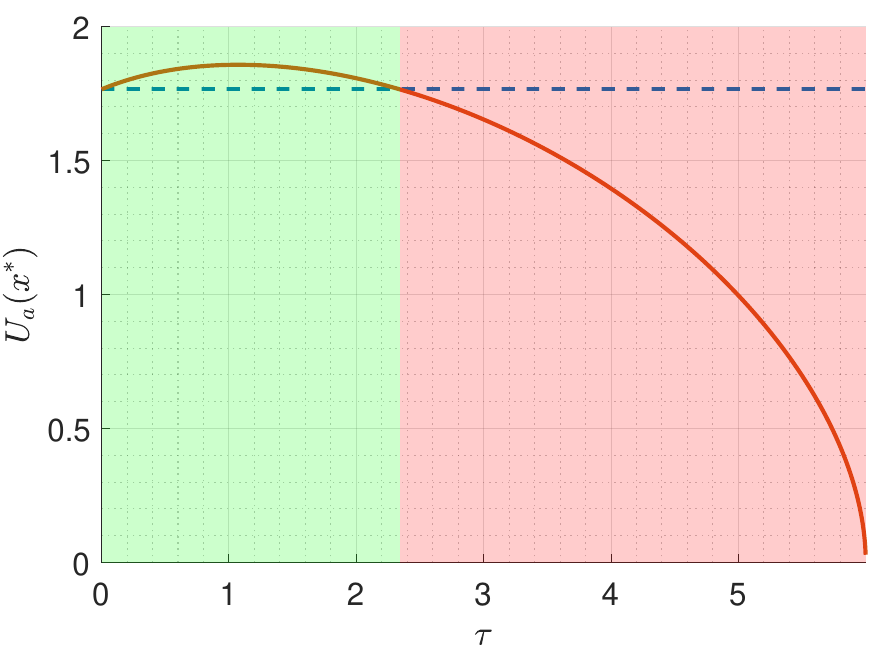}} \hfil
    \subfloat[Player $3$ \label{fig:3node:b}]{\includegraphics[width=0.24\textwidth]{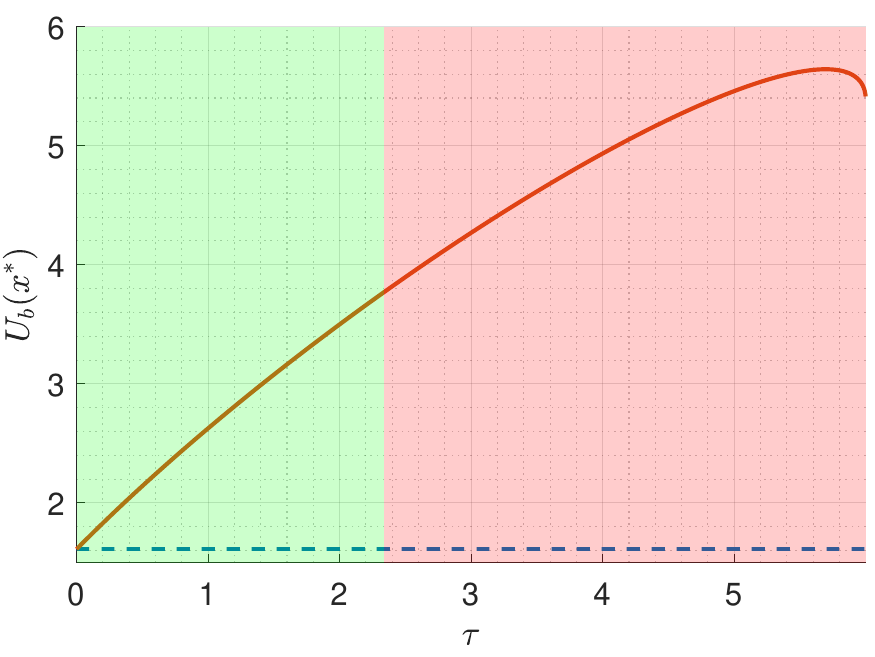}} 
    \caption{ Equilibrium payoffs as a function of the budgetary transfer $\tau$ for a $3$-line graph. The dashed line represent the equilibrium payoff with $\tau=0$. The region where a transfer is mutually beneficial is highlighted in green, while the red region highlights the values of $\tau$ where the transfer is not mutually beneficial.  }
    \label{fig:sims:3node}
\end{figure}

In Figure~\ref{fig:sims:3node} we show how the players' equilibrium payoff evolves with respect to $\tau$ for both the donating player $a$ and the receiving player $b$. Notice that the value of $U_a(x^*)$ and $U_b(x^*)$ increases for small values of $\tau$. This is equivalent to ensure that the rate of change of the equilibrium utilities are strictly positive with respect to $\tau$ when $\tau$ is close enough to $0$. Since the derivative of both players' utilities are positive we can ensure that there is some value of $\tau>0$ that improves both players' payoff, constituting a mutually beneficial coalition. 

\begin{figure}[htb!]
    \centering
    \includegraphics[width=0.2\textwidth]{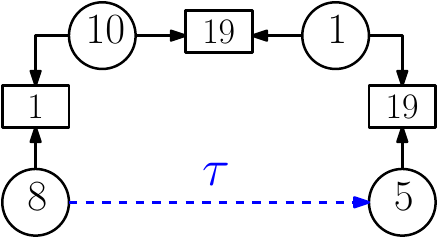}
    \caption{ Networked contest game in a $4$-line graph. The potential alliance is highlighted with the dashed arrow.   }
    \label{fig:4node}
\end{figure}

Similarly, we can analyze the scenario presented in Figure~\ref{fig:4node}. Here, we can analyze a donation from player $1$ to player $4$ in a $4$-node line graph. 

\begin{figure}[hbt]
    \centering
    \subfloat[Player $1$ \label{fig:4node:a}]{\includegraphics[width=0.24\textwidth]{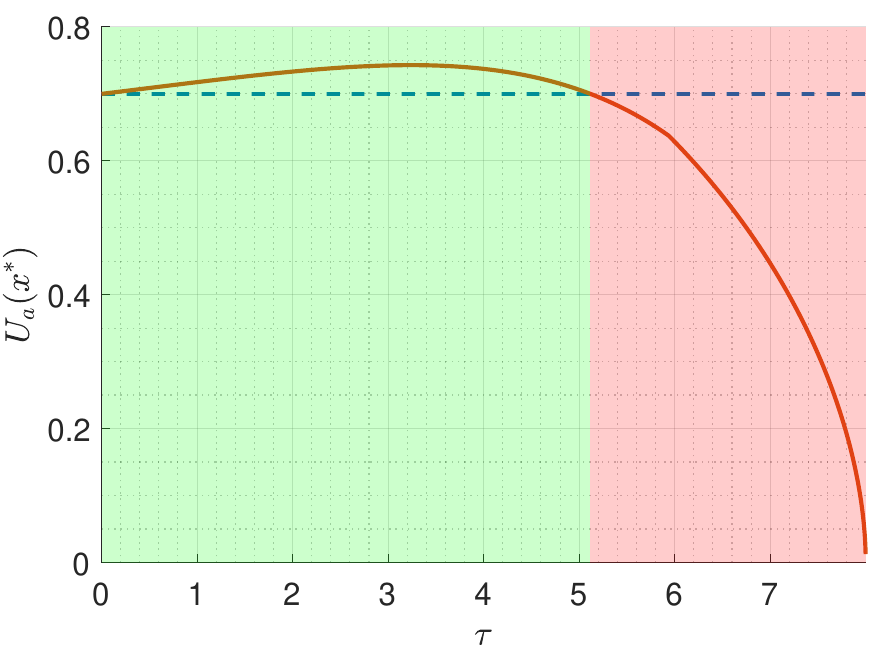}} \hfil
    \subfloat[Player $4$ \label{fig:4node:b}]{\includegraphics[width=0.24\textwidth]{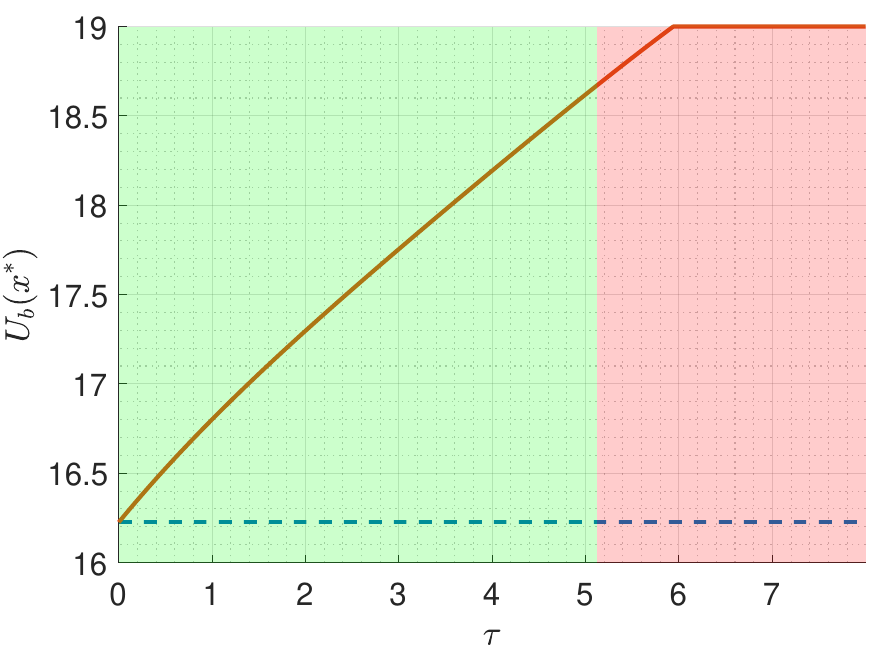}} 
    \caption{ Equilibrium payoffs as a function of the budgetary transfer $\tau$ for a $4$-line graph. The dashed line represent the equilibrium payoff with $\tau=0$. The region where a transfer is mutually beneficial is highlighted in green, while the red region highlights the values of $\tau$ where the transfer is not mutually beneficial. }
    \label{fig:sims:4node}
\end{figure}

Figure~\ref{fig:sims:4node} shows the equilibrium payoffs of the players as a function of the donation between the players $\tau$. Notice how the players' utilities increases for small values of $\tau$ as a consequence of their positive derivative at $\tau=0$. Therefore, the condition presented in Equation~\eqref{eq:cond} is a sufficient condition to guarantee the existence of a mutually beneficial budgetary transfer. 

\begin{figure}[htb!]
    \centering
    \includegraphics[width=0.45\textwidth]{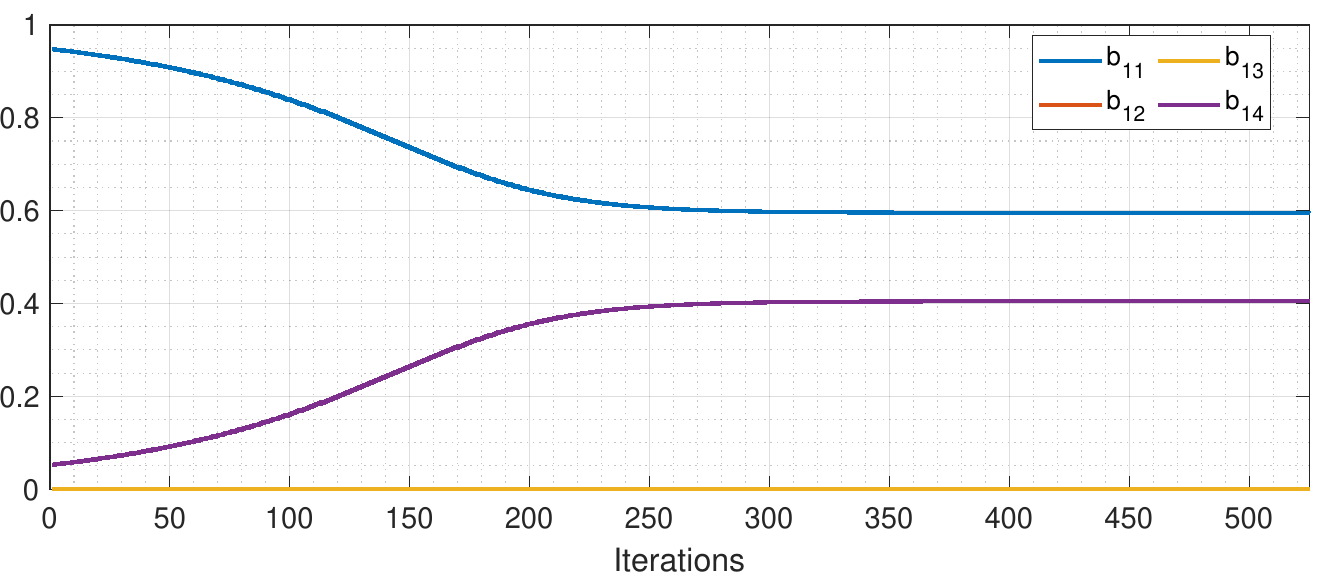}
    \caption{Gradient-based update rule to find the optimal donation from player $1$ to player $4$ in a $4$-line graph. }
    \label{fig:don4}
\end{figure}

Now, using the result in Section~\ref{sec:grad}, we aim to find the optimal transfer for the scenario presented in Figure~\ref{fig:4node}. With this in mind, we use the replicator dynamics to iteratively improve the payoffs for players $1$ and $4$~\cite{sandholm,discrep}. As a result, the vector of transfers $b_i$ evolve using the following update rule, 
\begin{equation} \label{eq:upd}
    b_{i,j} \gets \frac{\beta + f_{i,j}}{ \beta + \sum_j b_{i,j} f_{i,j} } b_{i,j},
\end{equation}
where $f_{i,j} = \frac{\partial U_i^*}{\partial b_{i,j}} $ represent how profitable if for player $i$ to donate to player $j$ and $\beta \in \R$ is a smoothing parameter that also ensure that $b_{i,j}$ still positive. The execution of the algorithm described in Equation~\ref{eq:upd} for the scenario depicted in Figure~\ref{fig:4node} is presented in Figure~\ref{fig:don4}. 

\begin{figure}[htb!]
    \centering
    \includegraphics[width=0.45\textwidth]{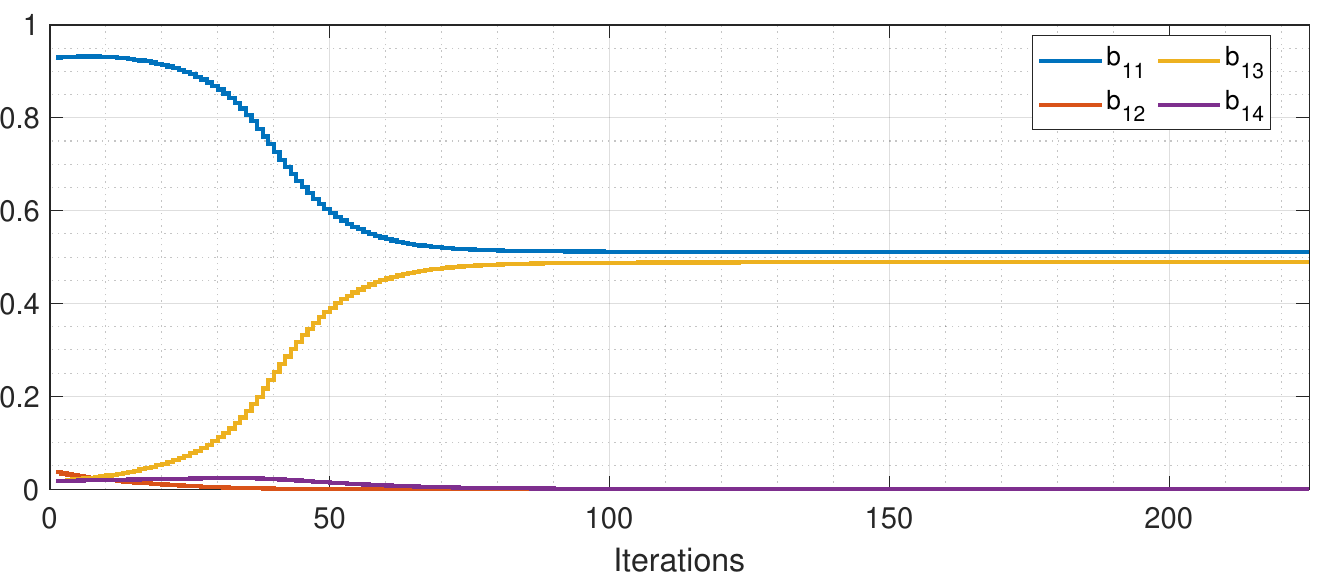}
    \caption{Gradient-based update rule to find the optimal donation from player $1$ to any player in a $4$-line graph. }
    \label{fig:donall}
\end{figure}

Notice that, in Figure~\ref{fig:don4}, the obtained optimal value of $b_{1,4}$ is approximately $0.4046$. This correspond to a donation $\tau \approx 3.2372$ which matches the plot in Figure~\ref{fig:4node:a}. Using the same update rule we can find the optimal donation when we give the option to player $1$ to donate to any other player in the network, i.e., $\D_1 = \crl{1,2,3,4}=\P$. The execution of the algorithm under this new scenario is presented in Figure~\ref{fig:donall}.

Unlike the scenario presented in Figure~\ref{fig:4node}, when the option player $1$ only can donate to player $3$, it chooses player $3$ to make an alliance. In Figure~\ref{fig:donall} can be noticed that the optimal value of $b_{1,3}$ is approximately $0.4885$, which correspond to a donation $\tau \approx 3.9081$. The equilibrium payoffs for the aforementioned scenarios are presented in Table~\ref{tb:eqpay}. Here, it can be verified that, in both scenarios, alliances are mutually beneficial. However, the alliance between players $1$ and $3$ is more profitable for player $1$. 

\begin{table}[htb]
    \caption{Equilibrum payoffs for optimally desgined donations}
    \centering
    \begin{tabular}{|c|c|c|c|c|}
    \hline
        Scenario                                        & $U_1^*$           & $U_2^*$   & $U_3^*$           & $U_4^*$           \\ \hline
        Without donation                                & 0.6995            & 18.8902   & 3.1852            & 16.2251           \\ \hline
        \makecell{ Optimal Donation \\ to Player $4$ }  & \textbf{0.7424}   & 16.2366   & 2.1613            & \textbf{27.8596}  \\ \hline
        \makecell{ Optimal Donation \\ to any player }  & \textbf{0.9831}   & 16.2759   & \textbf{10.2036}  & 11.5374           \\ \hline
    \end{tabular}
    \label{tb:eqpay}
\end{table}

\begin{figure}[htb!]
    \centering
    \includegraphics[width=0.25\textwidth]{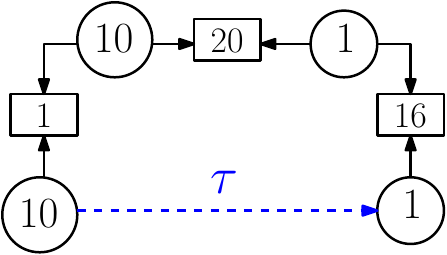}
    \caption{ Networked contest game in a $4$-line graph. The potential alliance is highlighted with the dashed arrow.   }
    \label{fig:var4}
\end{figure}

Finally, we present an instance of a mutually beneficial transfer in a $4$-line graph as shown in Figure~\ref{fig:var4}. 

\begin{figure}[htb!]
    \centering
    \subfloat[Player $1$ \label{fig:4node:vara}]{\includegraphics[width=0.24\textwidth]{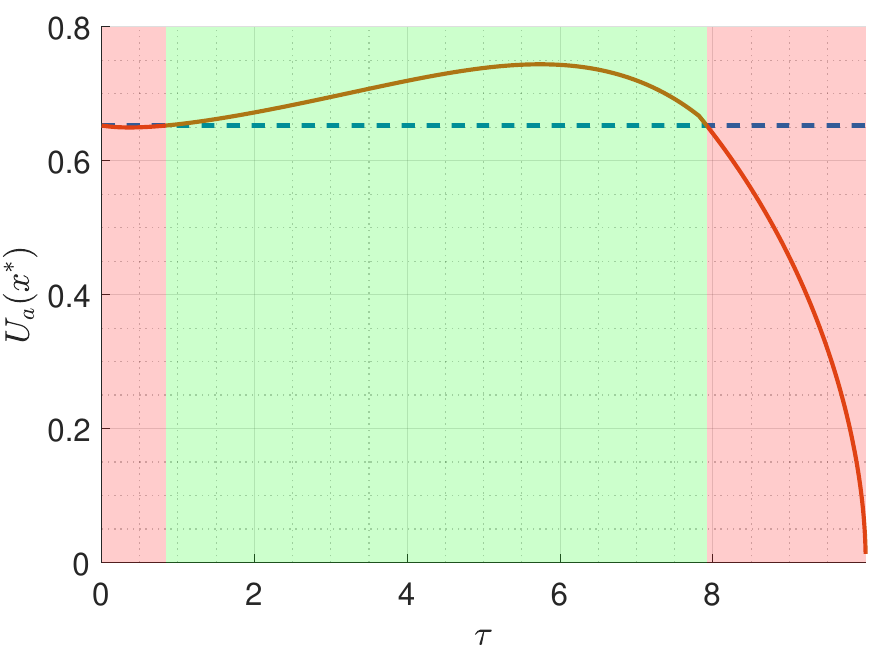}} \hfil
    \subfloat[Player $4$ \label{fig:4node:varb}]{\includegraphics[width=0.24\textwidth]{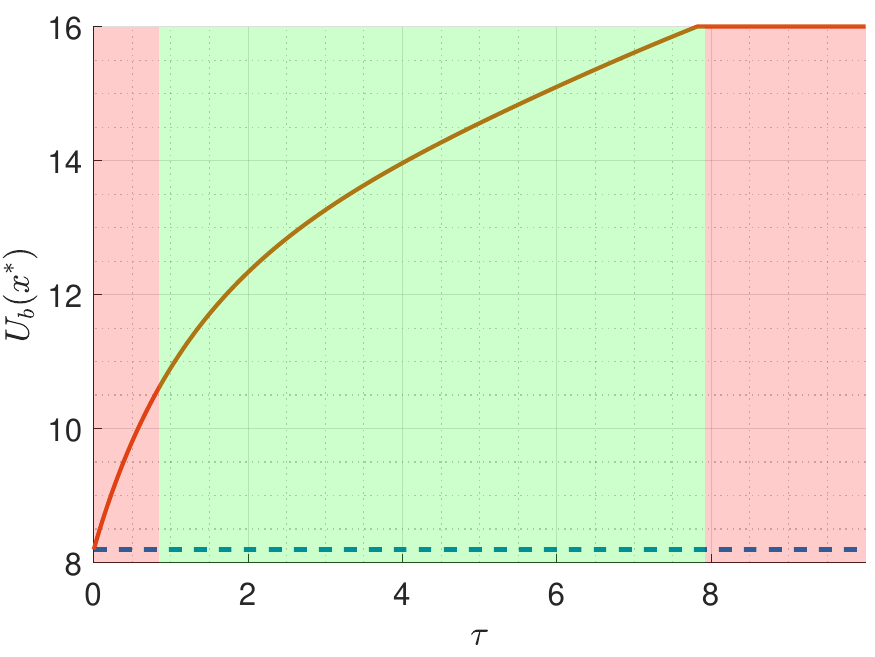}} 
    \caption{ Equilibrium payoffs as a function of the budgetary transfer $\tau$ for a $4$-line graph. The dashed line represent the equilibrium payoff with $\tau=0$. The region where a transfer is mutually beneficial is highlighted in green, while the red region highlights the values of $\tau$ where the transfer is not mutually beneficial. }
    \label{fig:sims:var}
\end{figure}

In Figure~\ref{fig:sims:var} can be noticed that there are values of $\tau>0$ where the budgetary transfer is mutually beneficial; however, there are not necessarily around $\tau=0$. Therefore, condition stated in Lemma~\ref{lm:cond} is sufficient but not necessary to guarantee the existence of a mutually beneficial transfer. 

While we are able to build instances for mutually beneficial transfers independently of the conflict graph, the generated instances always assume that the equilibrium payoffs increase around $\tau=0$. Thus, instances like the one presented in Figure~\ref{fig:var4} give us evidence that the parameter space for beneficial alliances is richer than the one used in our existence results. 

\section{Conclusions}\label{sec:conc}
In this work, we analyzed the effect of alliances with budgetary transfers in networked contest games. Under a different parametrization of the game, we were able to characterize the equilibrium payoffs for the players and check if a potential transfer between players is mutually beneficial for them. Using those conditions, we were able to build instances where a mutually beneficial alliance exist between two players under connectivity conditions on the conflict graph. With them, we were able to ensure that the existence of mutually beneficial alliances is independent to network structure that supports the contest game. However, its existence is tied to the budgets of the players and the values of the items they are fighting for. Therefore, for every player in a networked contest game, there exist potential strategic opportunities to form alliances with players that are significantly beyond its local network. 

Along our discussion we identified some future directions that worth a further analysis. While we were able to provide an expression for the gradient of the equilibrium utilities as a function of the donations over the network, there still a need to offer algorithmic guarantees for optimally designed donations. Moreover, we identify the existence of instances with mutually beneficial alliances that not satisfy the conditions presented in this work. Even if the existence of a beneficial alliance is guaranteed regardless the conflict network, a full characterization of mutually beneficial alliances requires further analysis. 

Additional future directions of this work include considering more general model of budgetary alliances. For instance, a cost associated to transferring resources could be taken into account. Additionally, we can consider that the effectiveness of resources after a transfer can be also affected. 

\section*{References}
\bibliographystyle{IEEEtran}
\bibliography{bib}

\begin{thebibliography}{10}
\providecommand{\url}[1]{#1}
\csname url@samestyle\endcsname
\providecommand{\newblock}{\relax}
\providecommand{\bibinfo}[2]{#2}
\providecommand{\BIBentrySTDinterwordspacing}{\spaceskip=0pt\relax}
\providecommand{\BIBentryALTinterwordstretchfactor}{4}
\providecommand{\BIBentryALTinterwordspacing}{\spaceskip=\fontdimen2\font plus
\BIBentryALTinterwordstretchfactor\fontdimen3\font minus \fontdimen4\font\relax}
\providecommand{\BIBforeignlanguage}[2]{{%
\expandafter\ifx\csname l@#1\endcsname\relax
\typeout{** WARNING: IEEEtran.bst: No hyphenation pattern has been}%
\typeout{** loaded for the language `#1'. Using the pattern for}%
\typeout{** the default language instead.}%
\else
\language=\csname l@#1\endcsname
\fi
#2}}
\providecommand{\BIBdecl}{\relax}
\BIBdecl

\bibitem{donation:conf}
G.~D{\'\i}az-Garcia, F.~Bullo, and J.~R. Marden, ``Beyond the `enemy-of-my-enemy' alliances: Coalitions in networked contest games,'' in \emph{IEEE Conference on Decision and Control}, 2023, pp. 2220--2225.

\bibitem{contth}
L.~C. Corch{\'o}n, M.~Serena \emph{et~al.}, ``Contest theory,'' \emph{Handbook of Game Theory and Industrial Organization}, vol.~2, pp. 125--146, 2018.

\bibitem{conttheory}
M.~Vojnovi{\'c}, \emph{Contest Theory: Incentive Mechanisms and Ranking Methods}.\hskip 1em plus 0.5em minus 0.4em\relax Cambridge University Press, 2015.

\bibitem{games:sec}
M.~Tambe, \emph{{Security and Game Theory: Algorithms, Deployed Systems, Lessons Learned}}.\hskip 1em plus 0.5em minus 0.4em\relax Cambridge University Press, 2011.

\bibitem{games:poac}
R.~Yang, B.~J. Ford, M.~Tambe, and A.~Lemieux, ``{Adaptive resource allocation for wildlife protection against illegal poachers},'' in \emph{Procedings of International Conference on Autonomous Agents and Multiagent Systems (AAMAS)}, 2014, pp. 453--460.

\bibitem{games:sensor}
A.~Ferdowsi, W.~Saad, and N.~B. Mandayam, ``Colonel {Blotto} game for sensor protection in interdependent critical infrastructure,'' \emph{IEEE Internet of Things Journal}, vol.~8, no.~4, pp. 2857--2874, 2020.

\bibitem{game:robots}
R.~Cui, J.~Guo, and B.~Gao, ``Game theory-based negotiation for multiple robots task allocation,'' \emph{Robotica}, vol.~31, no.~6, pp. 923--934, 2013.

\bibitem{survey}
Q.~Fu and Z.~Wu, ``Contests: Theory and topics,'' in \emph{Oxford Research Encyclopedia of Economics and Finance}, 2019.

\bibitem{borel}
E.~Borel, ``La th{\'e}orie du jeu et les {\'e}quations int{\'e}grales a noyau sym{\'e}trique,'' \emph{Comptes rendus de l’Acad{\'e}mie des Sciences}, vol. 173, no. 1304-1308, p.~58, 1921.

\bibitem{gw:blotto}
O.~Gross and R.~Wagner, ``A continuous {Colonel} {Blotto} game,'' Rand Project Air Force Santa Monica CA, Tech. Rep., 1950.

\bibitem{csf}
S.~Skaperdas, ``Contest success functions,'' \emph{Economic Theory}, vol.~7, pp. 283--290, 1996.

\bibitem{blotto:eq1}
M.~R. Baye, D.~Kovenock, and C.~G. De~Vries, ``The all-pay auction with complete information,'' \emph{Economic Theory}, vol.~8, pp. 291--305, 1996.

\bibitem{ads:cont}
L.~Friedman, ``Game-theory models in the allocation of advertising expenditures,'' \emph{Operations Research}, vol.~6, no.~5, pp. 699--709, 1958.

\bibitem{blotto:eq2}
B.~Roberson, ``The {Colonel} {Blotto} game,'' \emph{Economic Theory}, vol.~29, no.~1, pp. 1--24, 2006.

\bibitem{multi:cont}
A.~R. Robson \emph{et~al.}, ``{Multi-item Contests},'' 2005.

\bibitem{net:targets}
A.~Aghajan, K.~Paarporn, and J.~R. Marden, ``{A General Lotto Game over Networked Targets},'' in \emph{IEEE Conference on Decision and Control}, 2022, pp. 5974--5979.

\bibitem{net:conf}
S.~Cortes-Corrales and P.~M. Gorny, ``Generalising conflict networks,'' 2018.

\bibitem{reveal:1}
R.~Chandan, K.~Paarporn, and J.~R. Marden, ``{When showing your hand pays off: Announcing strategic intentions in Colonel Blotto games},'' in \emph{American Control Conference}, 2020, pp. 4632--4637.

\bibitem{reveal:2}
A.~Gupta, G.~Schwartz, C.~Langbort, S.~S. Sastry, and T.~Ba{\c{s}}ar, ``{A three-stage Colonel Blotto game with applications to cyberphysical security},'' in \emph{American Control Conference}, 2014, pp. 3820--3825.

\bibitem{division}
K.~Paarporn, R.~Chandan, M.~Alizadeh, and J.~R. Marden, ``{The division of assets in multiagent systems: A case study in team Blotto games},'' in \emph{IEEE Conference on Decision and Control}, 2021, pp. 1663--1668.

\bibitem{transf:1}
D.~Kovenock and B.~Roberson, ``{Coalitional Colonel Blotto games with application to the economics of alliances},'' \emph{Journal of Public Economic Theory}, vol.~14, no.~4, pp. 653--676, 2012.

\bibitem{transf:2}
D.~Rietzke and B.~Roberson, ``The robustness of ‘enemy-of-my-enemy-is-my-friend’alliances,'' \emph{Social Choice and Welfare}, vol.~40, no.~4, pp. 937--956, 2013.

\bibitem{sandholm}
W.~H. Sandholm, \emph{Population Games and Evolutionary Dynamics}.\hskip 1em plus 0.5em minus 0.4em\relax MIT Press, 2010.

\bibitem{discrep}
P.~D. Taylor and L.~B. Jonker, ``Evolutionary stable strategies and game dynamics,'' \emph{Mathematical Biosciences}, vol.~40, no. 1-2, pp. 145--156, 1978.

\end{thebibliography}

\appendix
\subsection{ Proof of Proposition~\ref{prop:3node} }
First, note that players $1$ and $3$ only have the option of allocating all their resources in the item they are competing for. Then, player $2$'s payoff is
\begin{align*}
    U_2(x)  &= v_1 \frac{x_1}{B_1-\tau + x_1} + v_2 \frac{x_2}{B_3 + \tau + x_2}, \\
            &= v_1 \frac{x_1}{B_1-\tau + x_1} + v_2 \frac{B_2 - x_1}{B_3 + \tau + B_2 - x_1}.
\end{align*}

Solving for optimality conditions, 
\begin{equation*} \resizebox{0.9\hsize}{!}{$ \begin{aligned}
    &\frac{\partial U_2(x)}{\partial x_1} = 0, \\
    &\iff v_1 \frac{B_1 - \tau}{\prth{B_1-\tau + x_1}^2} - v_2 \frac{B_3 + \tau}{\prth{B_3 + \tau + B_2 - x_1}^2} = 0, \\
    &\iff x_1^*(\tau) = \frac{ \prth{B_2+B_3+\tau} \sqrt{v_1}\sqrt{B_1-\tau} - \prth{B_1-\tau} \sqrt{v_2}\sqrt{B_3+\tau} }{ \sqrt{v_1}\sqrt{B_1-\tau} + \sqrt{v_2}\sqrt{B_3+\tau} }.
\end{aligned} $} \end{equation*}

Since $x_2 = B_2 - x_1$, 
\begin{equation*} \resizebox{0.9\hsize}{!}{$ \begin{aligned}
     x_2^*(\tau) = \frac{ \prth{B_1+B_2-\tau} \sqrt{v_2}\sqrt{B_3+\tau} - \prth{B_3+\tau} \sqrt{v_1}\sqrt{B_1-\tau} }{ \sqrt{v_1}\sqrt{B_1-\tau} + \sqrt{v_2}\sqrt{B_3+\tau} }.
\end{aligned} $} \end{equation*}

Now, note that we require that $x_1^*(0) > 0$ and $x_2^*(0) > 0$ to ensure that there is a mutually beneficial transfer. \begin{itemize}
    \item If $x_1^*(0) = 0$ then $U_1^*(0) = v_1 \geq U_1^*(\tau)$ for all $\tau \in [0,B_1)$. Thus, there is no value of $\tau$ that makes a beneficial alliance for player $1$. 
    \item If $x_2^*(0) = 0$ then $U_3^*(0) = v_2 \geq U_3^*(\tau)$ for all $\tau \in [0,B_1)$. Thus, there is no value of $\tau$ that makes a beneficial alliance for player $3$. 
\end{itemize}

Therefore, we require that $x_1^*(0) \in (0,1)$, 
\begin{align*}
    & x_1^*(0) = \frac{ (B_2+B_3)\sqrt{v_1B_1} - B_1\sqrt{v_2B_3} }{ \sqrt{v_1B_1} + \sqrt{v_2B_3} } > 0, \\
    &\iff \sqrt{ \frac{v_1 B_1}{ v_2 B_3 } } > \frac{B_1}{B_2+B_3},
\end{align*}
and, 
\begin{align*}
    & x_2^*(0) = \frac{ (B_1+B_2)\sqrt{v_2B_3} - B_3\sqrt{v_1B_1} }{ \sqrt{v_1B_1} + \sqrt{v_2B_3} } > 0, \\
    &\iff \frac{B_1+B_2}{B_3} > \sqrt{ \frac{v_1 B_1}{ v_2 B_3 } }.
\end{align*}

Combining the two inequalities into one give us, 
\begin{align*}
    & \frac{B_1+B_2}{B_3} > \sqrt{ \frac{v_1 B_1}{ v_2 B_3 } } > \frac{B_1}{B_2+B_3}, \\
    &\iff \frac{(B_1 + B_2)^2}{B_1B_3} > \frac{v_1}{v_2} > \frac{B_1 B_3}{(B_2 + B_3)^2},
\end{align*}
which matches the first condition. Now, finding the equilibrium payoff for player $1$ we obtain, 
\begin{align*}
    U_1^*(\tau) &= v_1 \frac{ B_1 - \tau }{ B_1 - \tau + x_1^*(\tau) },  \\
                &= \frac{ \sqrt{v_1} \sqrt{B_1-\tau} \prth{ \sqrt{v_1}\sqrt{B_1-\tau} + \sqrt{v_2}\sqrt{B_3+\tau} } }{ B_1 + B_2 + B_3 }.
\end{align*}

Similarly, for player $3$, 
\begin{align*}
    U_3^*(\tau) &= v_2 \frac{ B_3 + \tau }{ B_3 + \tau + x_2^*(\tau) },  \\
                &= \frac{ \sqrt{v_2} \sqrt{B_3+\tau} \prth{ \sqrt{v_1}\sqrt{B_1-\tau} + \sqrt{v_2}\sqrt{B_3+\tau} } }{ B_1 + B_2 + B_3 }.
\end{align*}

Now, we find the conditions on $\tau$ to ensure the existence of a mutually beneficial transfer. For player $1$, 
\begin{align*}
    & U_1^*(\tau) > U_1^*(0), \\
    &\iff \sqr{ 1 + \frac{v_1}{v_2} }^{-1} \sqr{ \prth{B_1 - B_3} - 2\sqrt{ \frac{v_1}{v_2} B_1 B_3 } } > \tau,
\end{align*}
and for player $3$, 
\begin{align*}
    & U_3^*(\tau) > U_3^*(0), \\
    &\iff \sqr{ 1 + \frac{v_2}{v_1} }^{-1} \sqr{ \prth{B_1 - B_3} + 2\sqrt{ \frac{v_2}{v_1} B_1 B_3 } } > \tau.
\end{align*}

Therefore, we can guarantee the existence of a $\tau > 0$ if, 
\begin{align*}
    B_1 - B_3 > 2\sqrt{ \frac{v_1}{v_2} B_1 B_3 },
\end{align*}
which matches the second condition.


\subsection{Proof of Lemma~\ref{lm:eqCost} }
For any $(i,j) \in \E$ note that the optimality conditions $\frac{\partial \hat{U_i}}{\partial x_{i,j}} = 0$ and $\frac{\partial \hat{U_j}}{\partial x_{j,i}} = 0$ imply that, \[ \frac{v_{i,j}}{\lambda_i} x_{j,i}^* = \prth{ x_{i,j}^* + x_{j,i}^* }^2 = \frac{v_{i,j}}{\lambda_j} x_{i,j}^*. \]

Therefore, solving for $x_{i,j}^*$ we obtain, \[ x_{i,j}^* = v_{i,j} \frac{\lambda_j}{ \prth{ \lambda_i + \lambda_j }^2 }, \] for any $(i,j)\in\E$. Replacing the definition of $x_{i,j}^*$ in Equation~\eqref{eq:ui} and the budget constraint we obtain,
\begin{align*} 
    U_i(x^*)    = \sum_{j\in\N_i} v_{i,j} \frac{ x_{i,j}^* }{ x_{i,j}^* + x_{ji}^* }    
                = \sum_{j\in\N_i} v_{i,j} \frac{ \lambda_j }{ \lambda_i + \lambda_j }.
\end{align*}
\begin{align*} 
    B_i     = \sum_{j\in\N_i} x_{i,j}^* 
            = \sum_{j\in\N_i} v_{i,j} \frac{\lambda_j}{ \prth{ \lambda_i + \lambda_j }^2 },
\end{align*}
which matches the expressions in Lemma~\ref{lm:eqCost}.
To establish the equivalence between the two parametrizations of the game we need to verify that the map described in Equation~\eqref{eq:Bi} is locally invertible in its domain. With that in mind, we build its Jacobian matrix $\J_{B/\lambda}(\lambda)$ as, 
\begin{align*}
    \sqr{ \J_{B/\lambda}(\lambda) }_{i,i} = \frac{\partial B_i}{\partial \lambda_i} 
    = -2 \sum_{j\in\N_i} v_{i,j} \frac{\lambda_j}{ \prth{ \lambda_i + \lambda_j }^3 }, 
\end{align*}
and, 
\begin{align*}
    \sqr{ \J_{B/\lambda}(\lambda) }_{i,j} = \frac{\partial B_i}{\partial \lambda_j} 
    =  v_{i,j} \frac{ \prth{ \lambda_i - \lambda_j } }{ \prth{ \lambda_i + \lambda_j }^3 } 
    = -\sqr{ \J_{B/\lambda}(\lambda) }_{ji}.
\end{align*}

Note that for any $q \in \R^n$, 
\begin{align*}
    q ^\top \J_{B/\lambda}(\lambda) q &= \frac12 q ^\top \prth{ \J_{B/\lambda}(\lambda) + \J^\top_{B/\lambda}(\lambda) } q \\
    &= q ^\top \diag\prth{ \sqr{ \J_{B/\lambda}(\lambda) }_{ii} } q < 0,
\end{align*}
for any $\lambda \in \R^n_{>0}$. Therefore, the Jacobian matrix is negative definite and non-singular for any $\lambda \in \R^n_{>0}$. By the inverse function theorem, we can ensure that $B : \R^n_{>0} \to \R^n_{>0}$ as defined by Equation~\eqref{eq:Bi} is locally invertible everywhere in its domain. Thus, the two game formulations are equivalent.


\subsection{ Proof of Lemma~\ref{lm:cond} }
Note that if $\frac{\partial U_i^*}{\partial \tau} > 0$ at $\tau=0$ then there exists a $\tau > 0$ that increases the value of $U_i^*$. Using the chain rule, 
\begin{equation} \label{eq:dUdt} \begin{aligned}
    \eval{ \frac{\partial U_i^*}{\partial \tau} }_{\tau=0} &= \eval{ \sum_k \frac{\partial U_i^*}{\partial \tilde{B_k} } }_{ \tilde{B_k} = B_k }  \eval{ \frac{\partial \tilde{B_k} }{\partial \tau} }_{\tau = 0}
    = \nabla_B^\top U_i^* \frac{\partial \tilde{B} }{\partial \tau}
\end{aligned} \end{equation}
with $\nabla_B U_i^*$ is the gradient of $U_i^*$ with respect to $B$. Using the definition of the Jacobian matrix we have, 
\begin{equation} \label{eq:dUdB} \begin{aligned}
    \nabla_{\lambda} U_i^* &= \J_{B/\lambda}^\top(\lambda) \nabla_{B} U_i^* \\ \iff \nabla_{B} U_i^* &= \sqr{ \J_{B/\lambda}^\top(\lambda) }^{-1} \nabla_{\lambda} U_i^*.
\end{aligned} \end{equation}

Therefore, using Equations~\eqref{eq:dUdt} and \eqref{eq:dUdB} we obtain, 
\[ \frac{\partial U_i^*}{\partial \tau} > 0 \iff \nabla_\lambda^\top U_i^* ~~ \J^{-1}_{B/\lambda}(\lambda) ~~ \frac{\partial \tilde{B} }{\partial \tau} > 0. \]
    
\subsection{ Structural Properties of Jacobians for Cycle Graphs }
During our discussion we exploit the equivalence between budget-constrained contest games and its per-unit-cost parametrization. This equivalence is captured through the relation presented in Equation~\eqref{eq:Bi}. In particular, we focus on the Jacobian of that map to analyze infinitesimal changes in the equilibrium payoff of the players $U_i^*$. For $n$-cycle graphs, the Jacobian is a $n\times n$ cyclic tridiagonal matrix, i.e., it can be written as, 
\begin{equation*} \resizebox{0.95\hsize}{!}{$ \begin{aligned}
    \J^{\top}_{B/\lambda}(\lambda) = \begin{bmatrix}
        a_1     & -b_1      & \cdots    & 0         & 0         & 0         & b_n       \\
        b_1     & a_2       & -b_2      & 0         & 0         & 0         & 0         \\
        \vdots  & b_2       & a_3       & -b_3      & \ddots    & 0         & 0         \\
        0       & \vdots    & b_3       & \ddots    & \ddots    & \vdots    & 0         \\
        0       & 0         & \ddots    & \ddots    & a_{n-2}   & -b_{n-2}  & \vdots    \\
        0       & 0         & 0         & \cdots    & b_{n-2}   & a_{n-1}   & -b_{n-1}  \\
        -b_n    & 0         & 0         & 0         & \cdots    & b_{n-1}   & a_n
    \end{bmatrix}
\end{aligned} $} \end{equation*}
where, 
\begin{align*}
    a_i     &= \frac{\partial B_i}{\partial \lambda_i} = -2\prth{ w_{i-1,i} \lambda_{i-1} + w_{i,i+1} \lambda_{i+1} } < 0, \\
    b_i     &= \frac{\partial B_i}{\partial \lambda_{i+1} } = w_{i,i+1} \prth{ \lambda_i - \lambda_{i+1} }, \\
    w_{i,j} &= \frac{v_{i,j}}{ \prth{\lambda_i + \lambda_j}^3 }.
\end{align*}

Line graphs are a particular case of cycle graphs where $w_{i,i+1}$ for only one $i$. For them, we can use the variable transform $K_{i,j} := w_{i,j}\lambda_i > 0$ with, 
\begin{align*}
    a_i     &= -2 \prth{ K_{i-1,i} + K_{i+1,i} }, \\
    b_i     &= K_{i,i+1} - K_{i+1,i}.
\end{align*}

Therefore, for simplicity, we use the variables $a_i$ and $b_i$ instead of the valuations $v_{i,j}$ and the costs $\lambda_i$ in some of our proofs. 

\subsection{ Proof of Lemma~\ref{lm:line} }
Let us assume that there is an instance for a $n-2$-node line graph where there exist a mutually beneficial transfer between two players. Without losing any generality, we assume that the players are labeled $\P=\crl{3,\cdots,n}$ and the transfer occurs from player $n-1$ to player $n$. Therefore, $v_{n-1,n} = 0$ and there are $n-3$ edges between the allied players. 

Since there is a mutually beneficial transfer, using Remark~\ref{rm:cond} we can ensure that $s_n > s_{n-1}$ and $t_n > t_{n-1}$ where, 
\begin{equation} \label{eq:lineq} \begin{aligned}
    \J^{\top}_{B/\lambda}(\lambda)~s = \beta_1 \nabla_\lambda U^*_{n-1}     &
    \text{ and } &
    \J^{\top}_{B/\lambda}(\lambda)~t = \beta_2 \nabla_\lambda U^*_{n},
\end{aligned} \end{equation} 
with $\beta_i > 0$ for $i=\crl{1,2}$. Note that, the gradients can be written as, 
\begin{equation*} \resizebox{0.95\hsize}{!}{$ \begin{aligned}
    & \beta_1 \nabla_\lambda U^*_{n-1}    = \beta_1 \begin{bmatrix} 0    & 0     & \cdots    & \frac{\partial U^*_{n-1}}{\partial \lambda_{n-2}}     & \frac{\partial U^*_{n-1}}{\partial \lambda_{n-1}}     & 0 \end{bmatrix}^\top,   \\
                                        &= \beta_1 \prth{ \lambda_{n-2} + \lambda_{n-1} } \begin{bmatrix} 0 & 0 & \cdots & K_{n_1,n-2} & -K_{n-2,n-1} & 0 \end{bmatrix}^\top, \\
                                        &=: \tilde{\beta}_1 \begin{bmatrix} 0 & 0 & \cdots & -\prth{2b_{n-2} + a_{n-1}} & a_{n-1} & 0 \end{bmatrix}^\top,
\end{aligned} $} \end{equation*}
and,
\begin{equation*} \resizebox{0.85\hsize}{!}{$ \begin{aligned}
    & \beta_2 \nabla_\lambda U^*_n      = \beta_2 \begin{bmatrix} \frac{\partial U^*_n}{\partial \lambda_1}   & 0     & \cdots    & 0 & 0 & \frac{\partial U^*_n}{\partial \lambda_n}  \end{bmatrix}^\top,   \\
                                        &= \beta_2 \prth{ \lambda_1 + \lambda_n } \begin{bmatrix} K_{n,1}   & 0     & \cdots    & 0 & 0 & -K_{1,n}  \end{bmatrix}^\top, \\
                                        &=: \tilde{\beta}_2 \begin{bmatrix} 2b_n - a_n   & 0     & \cdots    & 0 & 0 & a_n  \end{bmatrix}^\top,
\end{aligned} $} \end{equation*}
with $\tilde{\beta}_i > 0$ for $i=\crl{1,2}$. Thus, the two linear system of equations can be compressed into the $n-2\times n$ extended matrix, 
\begin{equation}\label{eq:extM1} \resizebox{0.9\hsize}{!}{$ \begin{aligned}
    & \sqr{  \begin{array}{c|cc} \J^{\top}_{B/\lambda}(\lambda) & \beta_1 \nabla_\lambda U^*_{n-1} & \beta_2 \nabla_\lambda U^*_n  \end{array} } = \\
    & \sqr{ \begin{array}{cccccc|cc}
        a_3     & -b_3      & \cdots    & 0         & 0         & b_n       & 0                                 & 2b_n - a_n    \\
        b_3     & a_4       & \cdots    & 0         & 0         & 0         & 0                                 & 0             \\
        \vdots  & \vdots    & \ddots    & \ddots    & \vdots    & 0         & \vdots                            & \vdots        \\
        0       & 0         & \ddots    & a_{n-2}   & -b_{n-2}  & \vdots    & -\prth{ 2 b_{n-2} + a_{n-1} }     & 0             \\
        0       & 0         & \cdots    & b_{n-2}   & a_{n-1}   & 0         & a_{n-1}                           & 0             \\
        -b_n    & 0         & 0         & \cdots    & 0         & a_n       & 0                                 & a_n
    \end{array} }
\end{aligned} $} \end{equation}
with $\tilde{\beta}_i = 1$ for $i=\crl{1,2}$. Note that we care about the solution of the last two rows of the extended matrix in Equation~\eqref{eq:extM1}. The $n\times n+2$ extended matrix, 
\begin{equation}\label{eq:extM2} \resizebox{0.9\hsize}{!}{$ \begin{aligned}
    \sqr{ \begin{array}{ccccccc|cc}
        a_1                                     & -b_1                      & 0                         & \cdots    & 0         & 0         & \frac{a_1}{b_1} \frac{a_2}{b_2} b_n   & 0                             & \frac{a_1}{b_1} \frac{a_2}{b_2} \prth{2b_n - a_n} \\
        b_1                                     & a_2 - \frac{b_1^2}{a_1}   & -b_2                      & \cdots    & 0         & 0         & 0                                     & 0                             & 0                                                 \\
        0                                       & b_2                       & a_3 - \frac{b_2^2}{a_2}   & \cdots    & 0         & 0         & 0                                     & 0                             & 0                                                 \\
        \vdots                                  & \vdots                    & \ddots                    & \ddots    & \ddots    & \vdots    & \vdots                                & \vdots                        & \vdots                                            \\
        0                                       & 0                         & 0                         & \ddots    & a_{n-2}   & -b_{n-2}  & \vdots                                & -\prth{2 b_{n-2} + a_{n-1} }  & 0                                                 \\
        0                                       & 0                         & 0                         & \cdots    & b_{n-2}   & a_{n-1}   & 0                                     & a_{n-1}                       & 0                                                 \\
        -\frac{a_1}{b_1} \frac{a_2}{b_2} b_n    & 0                         & 0                         & 0         & \cdots    & 0         & a_n + a_2 \frac{b_n^2}{b_2^2}\prth{ 1 - \frac{a_1a_2}{b_1^2} }        & 0 &  a_n + a_2 \frac{b_n^2}{b_2^2}\prth{ 1 - \frac{a_1a_2}{b_1^2} } \prth{ 2b_n - a_n }
    \end{array} },
\end{aligned} $} \end{equation}
has the same solution for the last $n-2$ rows as the extended matrix shown in Equation~\eqref{eq:extM1}. Moreover, under some conditions, the extended matrix in Equation~\eqref{eq:extM2} describe an instance for a $n$-node line graph, i.e., it can be written as, 
\begin{equation*}
    \sqr{ \begin{array}{c|cc} \J^{\top}_{\bar{B}/\bar{\lambda}}(\bar{\lambda}) & \alpha_1 \nabla_\lambda \bar{U}^*_{n-1} & \alpha_2 \nabla_\lambda \bar{U}^*_n  \end{array} }.
\end{equation*}

To preserve the negativity of the diagonal elements we require, 
\begin{equation*}
    \frac{a_1a_2}{b_1^2} =: \eta_1 \in (0,1).
\end{equation*}
In addition, let us define, 
\begin{equation*} 
    \frac{a_1}{a_3 \eta_1} \frac{b_2^2}{b_1^2} =: \eta_2.
\end{equation*} 
Thus, we can rewrite the extended matrix as, 
\begin{equation}\label{eq:extM3} \resizebox{0.9\hsize}{!}{$ \begin{aligned}
    \sqr{ \begin{array}{ccccccc|cc}
        \eta_1 \eta_2 \frac{b_1^2}{b_2^2} a_3   & -b_1                      & 0                         & \cdots    & 0         & 0         & \eta_1 \frac{b_1}{b_2} b_n            & 0                             & \eta_1 \frac{b_1}{b_2} \prth{2b_n - a_n} \\
        b_1                                     & \frac{\eta_1-1}{\eta_1\eta_2} \frac{b_2^2}{a_3}       & -b_2      & \cdots    & 0         & 0                                     & 0                             & 0      & 0 \\
        0                                       & b_2                       & \prth{1-\eta_2} a_3       & \cdots    & 0         & 0         & 0                                     & 0                             & 0                                                 \\
        \vdots                                  & \vdots                    & \ddots                    & \ddots    & \ddots    & \vdots    & \vdots                                & \vdots                        & \vdots                                            \\
        0                                       & 0                         & 0                         & \ddots    & a_{n-2}   & -b_{n-2}  & \vdots                                & -\prth{2 b_{n-2} + a_{n-1} }  & 0                                                 \\
        0                                       & 0                         & 0                         & \cdots    & b_{n-2}   & a_{n-1}   & 0                                     & a_{n-1}                       & 0                                                 \\
        -\eta_1 \frac{b_1}{b_2} b_n             & 0                         & 0                         & 0         & \cdots    & 0         & a_n - \frac{\eta_1-1}{\eta_2} \frac{b_n^2}{a_3}        & 0 &  a_n - \frac{\eta_1-1}{\eta_2} \frac{b_n}{a_3} \prth{ 2b_n - a_n }
    \end{array} },
\end{aligned} $} \end{equation}
which makes evident that $\eta_2 \in (0,1)$ and $\alpha_1 = 1$. Moreover, we require that $\alpha_2 \sqr{\J^{\top}_{\bar{B}/\bar{\lambda}}(\bar{\lambda})}_{n,n} = \sqr{\nabla_\lambda \bar{U}^*_n }_n$ with $\alpha_2 > 0$, i.e., 
\begin{align*}
    &\alpha_2 \sqr{ a_n - \frac{\eta_1-1}{\eta_2} \frac{b_n^2}{a_3} } = a_n - \frac{\eta_1-1}{\eta_2} \frac{b_n}{a_3} \prth{ 2b_n - a_n } \\
    \iff &\alpha_2 = \frac{ \eta_2 a_3 a_n - b_n(\eta_1-1)(2b_n - a_n) }{ \eta_2 a_3 a_n - (\eta_1 -1) b_n^2 }.
\end{align*}

Note that, for any $\eta_2 \in (0,1)$ and $\eta_1 = 1$, we have $\alpha_2 = 1$. Then, by continuity of $\alpha_2$ as a function of $\eta_1$ there must exist $\eps_\eta > 0$ such that $\alpha_2$ still positive for $\eta_1 = 1 - \eps_\eta$. The other equation that the extended matrix should satisfy is
\begin{align*}
    &\alpha_2 \crl{ 2 \sqr{\J^{\top}_{\bar{B}/\bar{\lambda}}(\bar{\lambda})}_{1,n} - \sqr{\J^{\top}_{\bar{B}/\bar{\lambda}}(\bar{\lambda})}_{n,n} } = \sqr{\nabla_\lambda \bar{U}^*_n }_1 \\
    \iff &\alpha_2 \crl{ 2 \eta_1 \frac{b_1}{b_2} b_n - \sqr{a_n - \frac{\eta_1-1}{\eta_2} \frac{b_n^2}{a_3}  } } = \eta_1 \frac{b_1}{b_2} \prth{2b_n - a_n} \\
    \iff &\frac{b_1}{b_2} = \frac{ \alpha_2 \sqr{a_n - \frac{\eta_1-1}{\eta_2} \frac{b_n^2}{a_3}  }  }{ \eta_1 \sqr{ 2 \alpha_2 b_n - \prth{2b_n - a_n} }  } =: \frac{1}{r_b}.
\end{align*}

With the definitions of $\eta_1$, $\eta_2$, $\alpha_2$ and $r_b$ we can define an instance with a mutually beneficial transfer for a $n$-node line graph from a instance for a $(n-2)$-node line graph. The steps can be enunciated as follows, \begin{enumerate}
    \item Pick $\eta_1 \in (0,1)$ and $\eta_2 \in (0,1)$ such that $\alpha_2 > 0$.
    \item Using the values of $\eta_1$, $\eta_2$ and $\alpha_2$ define $r_b$. 
    \item Pick a value of $b_1$. 
    \item Using the first $n$ columns of the extended matrix find the values of $a_i$ and $b_i$, or equivalently, the costs $\lambda_i$ and the values $v_{i,j}$. 
\end{enumerate}

\subsection{ Proof of Lemma~\ref{lm:cycle} }
Let us assume that the mutually beneficial transfer exists from node $1$ to node $n$ in a $n$-node line graph. This is equivalent to ensure that $\delta^\top s > 0$ and $\delta^\top t > 0$ with,
\begin{align*}
    \J^{\top}_{B/\lambda}(\lambda)~s &= \nabla_\lambda U^*_1, \\
    \J^{\top}_{B/\lambda}(\lambda)~t &= \nabla_\lambda U^*_n, \\
    \delta &= \begin{bmatrix} -1 & 0 & \cdots & 0 & 1 \end{bmatrix}^\top.
\end{align*}

For simplicity, let us write the systems of linear equations $\J^{\top}_{B/\lambda}(\lambda)~s = \nabla_\lambda U^*_1$ and $\J^{\top}_{B/\lambda}(\lambda)~t = \nabla_\lambda U^*_n$ as $Qs = \prth{\lambda_1 + \lambda_2}y$ and $Qt = \prth{\lambda_{n-1} + \lambda_n}z$ respectively. 

If we add an edge between nodes $1$ and $n$ with weight $v_{1,n}>0$ then we need to check that there still exists a mutually beneficial transfer, i.e., $\delta^\top \tilde{s} > 0$ and $\delta^\top \tilde{t} > 0$ with $\tilde{Q}\tilde{s} = \prth{ \lambda_1 + \lambda_2 } \tilde{y}$ and $\tilde{Q}\tilde{t} = \prth{ \lambda_{n-1} + \lambda_n } \tilde{z}$. 

Note that, $\tilde{y} = y + v_{1,n}\partial y$ and
\begin{align*}
    \tilde{Q} = Q + \partial Q = Q + v_{1,n} P^\top \begin{bmatrix} -2\lambda_n & \lambda_n-\lambda_1 \\ \lambda_1-\lambda_n & -2\lambda_1 \end{bmatrix} P
\end{align*}
with, 
\begin{align*}
    P = \begin{bmatrix}
        1 & 0 & \cdots & 0 & 0 \\
        0 & 0 & \cdots & 0 & 1
    \end{bmatrix},
\end{align*}
and
\begin{align*}
    \partial y = \begin{bmatrix}
        -\lambda_n \frac{\lambda_1 + \lambda_n}{\lambda_1 + \lambda_2}  & 0 & \cdots & 0 & \lambda_1 \frac{\lambda_1 + \lambda_n}{\lambda_1 + \lambda_2} 
    \end{bmatrix}^\top.
\end{align*}

For any factorization $\partial Q = v_{1,n} UV$ we have,
\begin{equation*} \resizebox{0.98\hsize}{!}{$ \begin{aligned}
    \prth{ Q + \partial Q }^{-1}    &= \prth{ Q + v_{1,n} UV }^{-1} \\
                                    &= Q^{-1} - v_{1,n} Q^{-1} U \prth{ I + v_{1,n} V Q^{-1} U }^{-1} V Q^{-1}.
\end{aligned} $} \end{equation*}

Therefore, 
\begin{equation*} \resizebox{0.98\hsize}{!}{$ \begin{aligned}
    \delta^\top \tilde{s} &= \delta^\top \tilde{Q}^{-1} \tilde{y} = \delta^\top \prth{ Q + \partial Q }^{-1} \prth{ y + v_{1,n}\partial y } \\
    &= \delta^\top Q y - v_{1,n} \delta^\top Q^{-1} U \prth{ I + v_{1,n} V Q^{-1}U  }^{-1} V Q^{-1}y \\
    &~~ + v_{1,n} \delta^\top Q^{-1}\partial y - v_{1,n}^2 \delta^\top Q^{-1} U \prth{ I + v_{1,n} V Q^{-1} U }^{-1} V Q^{-1} \partial y. 
\end{aligned} $} \end{equation*}

We can see the expression for $\delta^\top \tilde{s}$ as a continuous function of $v_{1,n}$ $f_y : [0,\infty) \to \R$ with $f_y(0) = \delta^\top Q y > 0$. Then, by continuity, it must exist $\bar{v}_y >0$ such that $f_y\prth{ v_{1,n} } > 0$ for all $v_{1,n} \in [0,\bar{v}_y)$. 

Since we can use a similar argument for $\delta^\top \tilde{t}$ then we can find a value for $v_{1,n}$ such that $\delta^\top \tilde{s} > 0$ and $\delta^\top \tilde{t} > 0$. Therefore, if there exist an instance for a mutually beneficial transfer in a $n$-node line graph then there exist at least one instance for a $n$-node cycle graph. 

\subsection{ Proof of Lemma~\ref{lm:neigh} }
For any set of budgets $B$ and values $v$ we can find the equilibrium allocation $x^*$. Since $\N_a = \crl{b}$ then $x_{a,b}^*= B_a$. Assume that there is a beneficial transfer $\tau>0$ from player $a$ to player $b$. Then, the budgets after the transfer are, 
\begin{align*}
    \tilde{B}_a &= B_a - \tau, \\
    \tilde{B}_b &= B_b + \tau, \\
    \tilde{B}_i &= B_i ~~~ \forall ~ i \in \P \setminus \crl{a,b}.
\end{align*}

Similarly, for the budgets $\tilde{B}$ and values $v$ we can define the equilibrium allocation $\tilde{x}^*$. The existence of a mutually beneficial transfer from $a$ to $b$ is equivalent to ensure that $U_a(x^*) < U_a(\tilde{x}^*)$ and $U_b(x^*) < U_b(\tilde{x}^*)$. Let us build the allocation $\hat{x}^*$ as follows, \begin{align*}
    \hat{x}_{a,b} &= B_a - \tau, \\
    \hat{x}_{b,a} &= \frac{B_a - \tau}{B_a} x_{b,a}^*, \\
    \hat{x}_{i,j} &= x_{i,j} \text{  otherwise }.
\end{align*}

Note that, by construction, $U_i(\hat{x}^*) = U_i(x^*)$ for all $i\in \P$. Then, $\hat{x}^*$ is also an equilibrium allocation since players are not incentivized to make unilateral changes. However, the budgets are defined as,  \begin{align*}
    \hat{B}_{i} &= B_i ~~~ \forall i \in \P\setminus\crl{a,b}, \\
    \hat{B}_{a} &= \hat{x}_{a,b}^* = B_a - \tau, \\
    \hat{B}_{b} &= \sum_{j \in \P} \hat{x}_{b,j}^* = \hat{x}_{b,a}^* + \sum_{j \in \P\setminus\crl{a}} \hat{x}_{b,j}^* \\
                &= \frac{B_a - \tau}{B_a} x_{b,a}^* + \sum_{j \in \P\setminus\crl{a}} x_{b,j}^*.
\end{align*}

Note that $\hat{B}_b < B_b < B_b + \tau$. Then, to get the budgets $\tilde{B}$ from the budgets $\hat{B}$ the player $b$ has to spend the additional budget $B_b + \tau - \hat{B}_b > 0$. Since all the marginal contributions $\frac{\partial U_b}{ \partial x_{b,j}}$ increase as $x_{b,j}$ increase then, for player $b$, all the allocations $\tilde{x}^*$ should be greater than $\hat{x}^*$, i.e., $\tilde{x}^*_{b,j} > \hat{x}^*_{b,j}$ for all $j \in \N_b$. In particular, \[ \tilde{x}_{b,a} > \hat{x}_{b,a} = \frac{B_a - \tau}{B_a} x^*_{b,a}. \]
This implies that $U_a(x^*) > U_a(\tilde{x}^*)$, which is a contradiction. Therefore, there cannot exist a mutually beneficial transfer from player $a$ to player $b$ if $\N_a=\crl{b}$. 

\subsection{ Proof of Proposition~\ref{prop:grad} }
Using the equality, 
\begin{equation*}
    \sum_{j\in\N_i}  v_{i,j} \frac{\lambda_j}{ \prth{ \lambda_i + \lambda_j }^2 } = \tilde{B}_i = \sum_{ j \in \D_i } b_{j,i}B_j ~~~~ \forall i,
\end{equation*}
and differentiating it with respect to $b_{k,l}$ we obtain, 
\begin{equation*} \resizebox{0.9\hsize}{!}{$ \begin{aligned}
    & \sum_j v_{i,j} \crl{ \frac{\partial}{\partial \lambda_i} \sqr{ \frac{\lambda_j}{ \prth{ \lambda_i + \lambda_j }^2 } } \frac{\partial \lambda_i}{ \partial b_{k,l} }  + \frac{\partial}{\partial \lambda_j} \sqr{ \frac{\lambda_j}{ \prth{ \lambda_i + \lambda_j }^2 } } \frac{\partial \lambda_j}{ \partial b_{k,l} } } = \sum_j \frac{\partial b_{j,i}}{\partial b_{k,l} } B_j \\
    &\iff \sqr{ -2 \sum_{j\neq i} v_{i,j} \frac{\lambda_j}{\prth{\lambda_i + \lambda_j}^3 } } \frac{\partial \lambda_i}{ \partial b_{k,l} } + \sum_{j\neq i} v_{i,j} \frac{ \prth{\lambda_i - \lambda_j} }{\prth{\lambda_i + \lambda_j}^3 } \frac{\partial \lambda_j}{ \partial b_{k,l} } = B_k \delta_{i,l} \\
    &\iff \inner{ \nabla_\lambda \tilde{B}_i , \frac{\partial \lambda}{\partial b_{k,l}} } = B_k \delta_{i,l} ~~~ \forall i
\end{aligned} $} \end{equation*}
where $\delta_{i,l} = 1 \iff i = l$. Stacking all the inner products into a single matrix multiplication we obtain, 
\begin{align*}
    \J^{\top}_{\tilde{B}/\lambda}(\lambda) \frac{\partial \lambda}{\partial b_{k,l} } = B_k e_l.
\end{align*}

Using the chain rule, we have that, 
\begin{align*}
    \frac{\partial U_i^*}{\partial b_{i,j}}     &= \prth{ \nabla_\lambda U_i^* }^\top \frac{\partial \lambda}{\partial b_{i,j} } \\
                                                &= B_i \prth{ \nabla_\lambda U_i^* }^\top \sqr{ \J^{\top}_{\tilde{B}/\lambda}(\lambda) }^{-1} e_j.
\end{align*}

Stacking the partial derivatives with respect to $b_{i,j}$ into the gradient we obtain, 
\begin{align*}
    \nabla_{b_i} U_i^* = B_i \sqr{ \J^{\top}_{\tilde{B}/\lambda}(\lambda) }^{-1} \nabla_\lambda U_i^*,
\end{align*}
with matches the expression in Equation~\eqref{eq:grad}.


\end{document}